\documentclass[11pt,prd,tightenlines,nofootinbib,superscriptaddress]{revtex4}

\usepackage[a4paper, centering, hmargin=2cm, vmargin=2.5cm]{geometry}
\usepackage{amsmath,amsfonts,amssymb,amsthm,mathrsfs,mathtools,bbm}
\usepackage[dvipsnames]{xcolor}
\usepackage{color, psfrag}
\usepackage[dvips]{graphicx}
\usepackage{braket}
\usepackage[linkbordercolor=cyan]{hyperref}
\usepackage{subcaption}
\DeclareMathOperator\arctanh{arctanh}

\DeclareMathOperator\chn{chn}

\newcommand{\C}{{\mathbbm C}}

\newcommand{\R}{{\mathbbm R}}

\newcommand{\cP}{{\mathcal P}}

\newcommand{\cT}{{\mathcal T}}

\newcommand{\cI}{{\mathcal I}}

\newcommand{\SU}{\mathrm{SU}}

\newcommand{\SL}{\mathrm{SL}}

\newcommand{\be}{\begin{equation}}
\newcommand{\ee}{\end{equation}}
\newcommand{\beq}{\begin{eqnarray}}
\newcommand{\eeq}{\end{eqnarray}}
\newcommand{\bes}{\begin{eqnarray}}
\newcommand{\ees}{\end{eqnarray}}
\newcommand{\mat} [2] {\left ( \begin{array}{#1}#2\end{array} \right ) }

\newcommand{\su}{{\mathfrak{su}}}

\newcommand{\so}{{\mathfrak{so}}}

\newcommand{\abs}[1]{\left|#1\right|}
\newcommand{\cobra}[1]{[ #1 |}
\newcommand{\coket}[1]{|#1]}

\newcommand{\Tr}{\text{Tr}}
\newcommand{\Pexp}{\cP\textrm{exp}}

\newcommand{\la}{\langle}
\newcommand{\ra}{\rangle}

\newcommand{\sgn}{\mathrm{sgn}}
\newcommand{\tr}{{\mathrm{Tr}}}
\newcommand{\f}{\frac}

\def\nn{\nonumber}
\def\pp{\partial}
\newcommand{\w}{\wedge}

\def\vphi{\varphi}
\def\eps{\epsilon}
\def\om{\omega}

\newcommand{\id}{\mathbbm{I}}

\def\rd{\mathrm{d}}
\def\sn{{\textrm{sn}}}

\newcommand{\bz}{\overline{z}}

\newcommand{\balpha}{\overline{\alpha}}
\newcommand{\bbeta}{\overline{\beta}}
\newcommand{\bE}{\overline{E}}
\newcommand{\bF}{\overline{F}}

\def\vX{\vec{X}}
\def\vY{\vec{Y}}

\def\vmu{\vec{\mu}}
\def\vnu{\vec{\nu}}
\def\vsigma{\vec{\sigma}}

\def\tcI{\tilde{\cI}}
\def\tvphi{\tilde{\vphi}}

\def\bfH{\mathbf{H}}
\def\bX{\mathbf{X}}

\def\ntau{u}

\begin{document}

\title{Elementary blocks of Loop Quantum Gravity}

\author{{\bf Mehdi Assanioussi}}\email{mehdi.assanioussi@ncbj.gov.pl}
\affiliation{National Centre for Nuclear Research,\\ Pasteura 7, 02-093 Warsaw, Poland}
\author{{\bf Etera R. Livine}}\email{etera.livine@ens-lyon.fr}
\affiliation{Univ Lyon, Ens de Lyon, Universit\'e Claude Bernard, CNRS,Laboratoire de Physique, F-69342 Lyon, France}

\begin{abstract}

We embark on the vast program of integrating the dynamics of Loop Quantum Gravity (LQG). Adopting the strategy of decomposing spin network states into small blocks of (quantum) geometry which can later be glued back together, we focus on the more modest objective of studying the Hamiltonian dynamics on the {\it candy graph}, that is two nodes linked together by an arbitrary number of edges and also having open edges. This elementary setting allows both for curvature to develop around the bulk loops and both non-trivial boundary data and dynamics on the open edges. We study this system at the classical level and leave the detailed of its quantum regime for future investigation. Working on a single loop with two external legs, we show how the LQG Hamiltonian ansatz reduces to a pair of non-linear differential equations, similar to the cubic Schr\"odinger equation, on the areas carried by the bulk links. We provide  analytical solutions to this evolution equation, identifying oscillatory modes (bounded modes) and divergent modes (similar to bouncing cosmological trajectories). This provides an explicit template for future investigations of LQG dynamics on more sophisticated spin network architecture built as arrays of candy graphs.

\end{abstract}

\maketitle
%%%%%%%%%%%%%%%%%%%%%%%%%%%%%%%%%%%%%%%%%%%%%%%%%%%%%%%%
{\small\tableofcontents}

\newpage

%%%%%%%%
\section*{Introduction}
%%%%%%%%

Loop Quantum Gravity (LQG) is one of the main approaches to quantum gravity \cite{Ashtekar:2004eh,Rovelli:2004tv,Thiemann:2007pyv}. It is a Hamiltonian framework for general relativity, based on the formulation of the theory \`a la Cartan in terms of transport and connection instead of distance and metric. The goal of any Hamiltonian approach to quantum gravity is to provide a consistent representation on suitable quantum states of geometry of the bulk Hamiltonian constraints and corresponding symmetries -space-time diffeomorphisms in particular- and of the algebra of conserved boundary charges. LQG looks for such a representation with quantum states defined as networks of entangled quanta of volumes interface with quanta of areas. These are called {\it spin networks}; they are the eigenstates of areas and volumes, with discrete spectra, and interpreted as fuzzy dynamical space-time lattices.

Since the theory's starting point is a new definition and description of the geometry at the Planck scale, the challenge is to understand the renormalization flow of LQG under coarse-graining from the Planck scale up to our scale and astrophysical scales, thereby shedding light on the emergence of the classicality of the space-time, of the continuum geometry and space-time diffeomorphisms (i.e. the equivalence and relativity principles).
There are proposals for the dynamics, either in the canonical setting of the original Hamiltonian formulation e.g.\cite{Thiemann:1996aw,Thiemann:1996av,Giesel:2006uj,Giesel:2006uk,Assanioussi:2015gka} or in the covariant setting of spinfoam path integrals e.g. \cite{Reisenberger:1996pu,Engle:2007wy,Rovelli:2014ssa}. But we still need to develop appropriate rigorous tools to extract, in a systematic and consistent way, the renormalization flow of the theory from the fundamental dynamics to effective modified gravity theories.
While most works in LQG nowadays are based, at least for the canonical formulation of the theory, on the kinematics (the consequences of a discrete spectrum for the area and volume operators, e.g. for black hole entropy) or on mini-superspace quantization (e.g. loop quantum cosmology and polymerized black holes), our take on this general endeavor is to proceed to a systematic in-depth analysis of the Hamiltonian dynamics, identify the basic dynamical building blocks of the theory, and investigate their various dynamical decoupled and coupled regimes. With the long-term ambitious objective of exploring and understanding the phase diagram of the possible Planck scale dynamics of spin networks.

Here, we take a first, modest but solid, step in this direction. We revisit the 2-vertex spin network configuration originally introduced in \cite{Borja:2010gn}, and make the crucial upgrade of allowing for a boundary. This yields  what we call the ``candy graph'', made of two nodes -representing elementary volume excitations- with an arbitrary number of links between them as well as open links attached to either node. This configuration allows for local curvature excitations around loops between the two nodes in the bulk of the candy graphs and for non-trivial boundary states on the open links.
Working with a boundary allows to study the effect of non-trivial boundary data of the bulk dynamics in LQG, which is crucial to investigate the holographic properties of the theory and, more generally the bulk-to-boundary propagation and vice-versa, but also to assemble candy graphs together to form larger more elaborate spin network architectures.

We would like to consider this candy graph as a basic template for LQG, the equivalent of the harmonic oscillator in quantum field theory, that the elementary building block of theory and a proof-of-concept for the basic dynamical mechanisms.
The present paper focuses on the classical dynamics of the model. We introduce a polynomial gauge-invariant Hamiltonian ansatz coming directly from the discretization of general relativity's Hamiltonian constraint. We exactly solve the model, and we identify the various regimes for a fixed geometry boundary conditions, putting in light bounded modes and accelerating modes. We highlight their relevance in the context of LQG's renormalization flow.  We discuss forced evolution for varying boundary conditions and possibilities for endowing the system with non-trivial boundary dynamics.
This is indeed a first step and we envision follow-up papers on the graph changing dynamics of the model upgraded with little loops attached to its nodes, on the quantum dynamics, focusing in particular on evolution of entanglement between quanta of volumes, and on the possibility of long-range propagation on scalable spin network architectures and potential simulations of the LQG dynamics on quantum circuits. \\

The paper is organized as follow. A first section is dedicated to an overview of the loop quantum gravity formalism, the definition of a regularized Hamiltonian acting on classical spin networks and the introduction of all the necessary notations and conventions. In the second section, we apply the formalism to the candy graph. We derive the explicit equations of motion for the area and holonomy variables of the models. Section III provides their analytical solutions and identifies the two types of modes - oscillatory modes, that ought to be considered as the bounded modes of the system, and accelerating modes, that grow and eventually diverge similarly to cosmological solutions. These are the template solutions, which we wish to generalize to more elaborate spin network states in the future.
Section IV proposes scenarios for evolving and dynamical boundary data, while the concluding outlook section discusses potential technical improvements and future development perspectives.

%%%%%%%%
\section{Loop quantum gravity}\label{LQG}
%%%%%%%%

Loop quantum gravity is a background independent approach to quantize general relativity based on the Ashtekar-Barbero Hamiltonian formulation.
The first canonical formulation of general relativity to be introduced is the ADM formulation \cite{Arnowitt:1959ah, Arnowitt:1962hi}, where the configuration variable is the spatial metric on the hypersurface. This formulation was and still is a very successful formalism in investigating classical solutions in general relativity. Unfortunately, the canonical quantization program cannot be completed in a rigorous and fully background independent manner due to various technical difficulties \cite{Thiemann:2007pyv}. Another issue in the ADM formulation resides in coupling gravity to Dirac spinors. Indeed, in order to couple Dirac spinors to gravity, and ultimately coupling fermions to quantum gravity, one must use the tetrad formulation of general relativity, also known as the tetrad Palatini formulation or the Palatini-Cartan formulation. 

In the Palatini-Cartan formulation, the tetrads $e_I^\mu$ and the spin connection $\omega_\mu^{IJ}$ (with $I,J$ being indices in the Lorentz algebra $\so(1,3)$ for Lorentzian spacetime and in the algebra $\so(4)$ for Euclidean spacetime) are the independent variables of the theory. The spacetime metric is recovered from the inverse of the tetrads (co-tetrads) as $g_{\mu\nu} = \eta_{IJ} e_\mu^I e_\nu^J$. When considering the equations of motion, the spin connection turns out to be entirely determined by the tetrads \cite{Ashtekar:2004eh}.  One can perform the Legendre transform to arrive at a Hamiltonian formulation of the theory. The result is a theory with second class constraints, which when solved simply lead to the ADM formulation expressed in terms of the triads $e_i^a$, which are the projections of the tetrads on the spatial slice with internal indices in the algebra $\so(3) \equiv \su(2)$, and with no reference to the original spin connection \cite{Ashtekar:1991hf}. The Hamiltonian formulation of the tetrad Palatini theory can be improved by introducing a new term in the action, such that the equations of motion are unaffected. This "topological" term is given by another invariant obtained from the tetrads and the spin connection, it is the so called Holst term \cite{Holst:1995pc}. 

The Palatini-Cartan-Holst action provides a description where the phase space of the theory is parametrized by the densitized co-triads $E_i^a \equiv \abs{\det(e)} e_i^a$ and an $\su(2)$-connection $A_a^i$, forming the Ashtekar-Barbero variables and satisfying the canonical Poisson brackets. The constraints obtained in this formulation \cite{Ashtekar:2004eh, Thiemann:2007pyv} consist of the Gauss constraint ${\mathcal G}(\Lambda)$,  the scalar constraint $H(N)$ and the spatial diffeomorphism (or vector) constraint $\mathcal{V}(\vec{N})$. They form a first class constraints algebra, but not a Lie algebra because it depends on structure functions on the phase space rather than structure constants. The spatial diffeomorphisms constraint $\mathcal{V}(\vec{N})$ and the scalar constraint $H(N)$ generate spatial diffeomorphisms and orthogonal diffeomorphisms (i.e.~time translations) respectively, while the Gauss constraint generates infinitesimal $SO(3)$ rotations and ensures the invariance of the spatial metric and its conjugate momentum under the action of the rotation group. Consequently, general relativity is cast in the form of an $SU(2)$-gauge theory, with spacetime diffeomorphisms symmetry. This brings general relativity closer to the formalism of standard gauge theories and allows to employ techniques used in the later to study gravitational physics. In particular, while the ADM formalism suggests to consider the phase space polarization in which the spatial metric is the configuration variable, and consequently the triad as the configuration variable in the Palatini-Cartan formulation, the Ashtekar-Barbero formulation obtained from the Palatini-Cartan-Holst action provides a polarization in line with the standard treatment of gauge theories, where the connection is taken to be the configuration variable. General relativity can then be formulated as a Yang-Mills-like theory, with simply a different Hamiltonian. This choice of phase space polarization is what opened the door to implement the quantization program of the theory, which eventually went beyond what was realized within quantum geometrodynamics and gave rise to loop quantum gravity.

Because one aims at a quantum theory of gravity, the quantization procedure must be consistent with background independence of the classical theory. In particular, background independence excludes working with distribution-valued quantum operators and consequently one cannot directly quantize the Ashtekar-Barbero variables $(A,E)$ since their Poisson brackets are singular. One is then compelled to use smeared phase space variables. Inspired by Yang–Mills theory, one requires that the smeared variables transform nicely under the gauge transformations, and is therefore led to consider Wilson loops as variables in the theory. Wilson loops are holonomies obtained by taking the path ordered exponential of the integral of the connection $A$ along closed embedded loops. Since the holonomies are defined through one dimensional smearing, it is geometrically natural that the conjugate variable would be defined through a co-dimension one (two dimensional) smearing of the densitized triads. This indeed gives the correct conjugate phase space variables, the so-called fluxes.

%%%%%%%%
\subsection{Holonomy-flux algebra and quantization}\label{I.A}
%%%%%%%%

Given oriented $1$-dimensional paths $\{e\}$ (edges) and $2$-dimensional surfaces $\{S\}$ embedded in the spatial hypersurface $\Sigma$, the holonomies $h_e[A]$ and fluxes $P_{S,\xi}$ are defined as:
\begin{align}
\label{def-holonomyflux}
	h_e[A] = \Pexp \left(-\int_e A \right) \ , \qquad
	P_{S,\xi} = \frac{1}{2}\int_S dx^b\wedge dx^c\epsilon_{abc}\xi^i(x) {E}^a_i(x)
\end{align}
where $A = A^i_a\tau_i \otimes dx^a$ is the $\su(2)$-valued connection $1$-form, $\tau_i$ being an orthonormal basis of the $\su(2)$ Lie algebra, $\Pexp$ denotes the path-ordered exponential and $\xi$ is a function valued in $\su(2)$.

Holonomies and fluxes still separate the phase space points identified by $A$ and $E$. They generate a Poisson algebra and transform nicely under the action of $SU(2)$-gauge transformations. Their Poisson bracket reflects the action of the fluxes on the holonomies as functional derivatives with respect to the connection variable. The Poisson bracket of two holonomies vanishes, while the one of two fluxes does not. The fluxes are derivative operators acting on the holonomies, hence the Lie algebra between holonomies and fluxes is given by the geometric Lie algebra of functions and vector fields on the configuration space. Since vector fields do not generally commute, we obtain that the Poisson bracket of fluxes does not vanish \cite{Ashtekar:2004eh}.

Given this holonomy-flux algebra, one proceeds with the quantization by introducing the space of cylindrical functions of the configuration variable $A$. These correspond to complex valued functions depending on the connection $A$ through finitely many holonomies:
\begin{equation}
	\Psi[A]\ =\ \psi(h_{e_1}[A],\ldots ,h_{e_n}[A]) 
\end{equation} 
with $\psi\ :\ {\rm SU}(2)^n \rightarrow \mathbbm{C}$. The set of embedded edges $\Gamma\coloneq\{e_1,...,e_n\}$ is called the graph of $\Psi$.
The space of all cylindrical functions with a graph $\Gamma$ is denoted ${{\rm Cyl}}_\Gamma$ and the space of all cylindrical functions is denoted ${\rm Cyl}\coloneq \cup_\Gamma\ {\rm Cyl}_\Gamma$. The kinematical Hilbert space of LQG, denoted ${\cal H}_{\rm kin}$, is then defined as the completion of ${\rm Cyl}$ with respect to the norm defined by a natural scalar product \cite{Ashtekar:1993wf, Ashtekar:1995zh, Lewandowski:2005jk} induced by the Haar measure on $SU(2)$. The Hilbert space ${\cal H}_{\rm kin}$ admits an orthonormal basis consisting of spin network functions. A spin network function is defined by an embedded graph $\Gamma$, with spins (labeling $SU(2)$ irreducible representations) assigned to the edges $\{e_i\}$ and $SU(2)$ tensors (intertwiners coupling the spins) assigned to the vertices $\{v_i\}$. The Hilbert space ${\cal H}_{\rm kin}$ is then expressed as the orthogonal sum ${\cal H}_{\rm kin} = \bigoplus_{\Gamma} {\cal H}_\Gamma$, where $\Gamma$ ranges over all the classes of embedded graphs and ${\cal H}_\Gamma$ is the Hilbert space defined as the completion of the space ${\rm Cyl}_\Gamma$ of cylindrical functions with the graph $\Gamma$. Each of these Hilbert spaces ${\cal H}_\Gamma$ is of the form $L^2 (SU(2)^{n_e}, d\mu_H)$, where $n_e$ is the number of edges of $\Gamma$, each edge carries one copy of $SU(2)$ and $d\mu_H$ is the product Haar measure on $SU(2)^{n_e}$.
As for the operator algebra acting in ${\cal H}_{\rm kin}$, every cylindrical function $\Psi$ (including single holonomies) defines a multiplication operator, while the quantum flux acts as a derivative operator.

The implementation and solving of the Gauss and spatial diffeomorphisms constraints are straightforwardly achieved via group averaging techniques \cite{Ashtekar:1995zh}. This provides two Hilbert spaces: the Hilbert space of $SU(2)$ gauge invariant states $\mathcal{H}_G \subset \mathcal{H}_\text{kin}$, and the Hilbert space of gauge and spatial diffeomorphisms invariant states $\mathcal{H}_\text{diff} \subset Cyl^*$, where $Cyl^*$ is the algebraic dual of the space $Cyl$. In contrast, since the scalar constraints do not generate a group, these cannot be implemented via group averaging techniques. In the canonical approach, one follows the Dirac prescription and attempts to define a quantum operator acting in $\mathcal{H}_G$ or $\mathcal{H}_\text{diff}$.
There have been several proposals to quantize the scalar constraint, and they primarily differ on how to regularize the curvature term $F_{ab}$ and the definition of the Lorentzian part. However, one can classify these different proposals within two categories: the graph changing operators \cite{Thiemann:1996aw, Assanioussi:2015gka} and the graph preserving ones \cite{Giesel:2006uj, Giesel:2006uk, Alesci:2014aza}, where the latter is inspired by the methods of lattice gauge theory. In the present article, we consider a specific graph preserving prescription to define the Hamiltonian, which we briefly review in what follows. We will not go as far as to define the Hamiltonian operator, as we are only interested in the regularized classical expression which we use to define the classical dynamics on the phase space of the theory with a fixed graph.

The scalar constraint in the Ashtekar-Barbero variables is given as 
\begin{align}\label{AB.Const.}
		H(N) = \frac{1}{2\kappa\gamma^2} \int_\Sigma d^3x\,N(x) \biggl( \frac{\epsilon_{ijk}E^a_i(x)E^b_j(x)F_{ab}^k(x)}{\sqrt{|\det E(x)|}} + \left(1-s \gamma^2\right) \sqrt{|\det E(x)|} \,R(x)\biggr)
\end{align}
where $F = dA + A\wedge A $ is the curvature $2$-form of the connection $A_a^i$, $R$ is the Ricci scalar of the metric tensor $q_{ab}$ on the $3$-dimensional spatial manifold, $\gamma$ is the Immirzi-Barbero parameter, while $N$ is the lapse function. Note that in the case of $s= 1$, namely Euclidean spacetime, the choice $\gamma =\pm 1$ eliminates the second term in the constraint and for that reason we call the first and second terms the Euclidean and Lorentzian parts respectively. The expression in \eqref{AB.Const.} is regularized via approximation of the integral over $\Sigma$ by a Riemannian sum over some partition $\mathscr{C}$ made of a collection of cells $\Delta$. The lapse $N(x)$ is evaluated at points $x_\Delta$ chosen in each cell, the densitized triads are replaced by fluxes, the curvature of the connection is replaced by holonomies along closed loops $\alpha_{IJ}(\Delta)$. Next, we adapt the partition to a chosen graph by matching the cells with the vertices $v$ and the splitting the boundary of each cell into surfaces such that each surface is dual to a single edge of the graph. The holonomies and fluxes are then restricted to the ones specified by the graph and its dual partition. The closed loops associated to the holonomy in the Hamiltonian are chosen to match the preexisting loops on the given graph, specifically the smallest closed loops (each defined in terms of the number of edges forming the loop) involving a pair of edges at a vertex. This leads to an action of the Hamiltonian which couples holonomies in a fixed representation (usually taken to be the fundamental representation, i.e.~$1/2$) to the ones preexisting in the state, without changing its graph. Hence, for a given graph $\Gamma$, we obtain an expression of the regularized Hamiltonian functional of the form
\begin{align}\label{LQG.Ham}
	H(N) &= \frac{1}{2\kappa\gamma^2} \sum \limits_{v \in \Gamma} N(v) \left( C_E(v) + (1- s\gamma^2) C_L(v) \right)
\end{align}
with the Euclidean part \cite{Assanioussi:2015gka} $C_E$ and the Lorentzian part \cite{Alesci:2014aza} $C_L$ given as
\begin{equation}\label{LQG.Ham.Parts}
	\begin{aligned}
		C_E(v) &= -\frac{(\kappa \gamma)^2}{2} k_E(v) \sum \limits_{I\neq J} \epsilon_{ijk} \text{Tr} \left[ (h_{\alpha_{IJ}} - h_{\alpha_{JI}}) \tau^i \right] P_{S_{e_I}}^j P_{S_{e_J}}^k V^{-1}_v \\[2ex]
		C_L(v) &= \frac{(\kappa \gamma)^2}{2} k_L(v)  \sum \limits_{I\neq J} \sqrt{ (\epsilon_{ijk} P_{S_{e_I}}^j P_{S_{e_J}}^k) (\epsilon_{i j' k'} P_{S_{e_I}}^{j'} P_{S_{e_J}}^{k'})} \left( \frac{2\pi}{\lambda_{IJ}} - \Theta_{IJ} \right) V^{-1}_v
	\end{aligned}
\end{equation}
where the sums are over pairs of edges not belonging to the same germ, $h_{\alpha_{IJ}}$ is the holonomy in the fundamental representation associated to the smallest closed loop $\alpha_{IJ}$ defined by the pair of edges $e_I$ and $e_J$ at the vertex $v$, and the fluxes $P_{S_{e_I}}^j\equiv P_{S_{e_I},\tau^j}$ are the smeared observables of the triad field as defined in \eqref{def-holonomyflux}.
The factors $k_E(v)$ and $k_E(v)$ are averaging coefficients which depend on the valence of the vertex, while $\lambda_{IJ}$ is a constant associated to the pair of edges $(I,J)$ in the graph and may vary from one pair to another. $\Theta_{IJ}$ is the regularized dihedral angle \cite{Major:1999mc, Alesci:2014aza} associated to a pair of edges $(e_I, e_J)$ at a vertex of the graph:
\begin{equation}
	\Theta_{e_I,e_J} \coloneq \pi - \arccos \left[ \frac{ P_{S_{e_I}}^i P_{S_{e_J}}^i}{\sqrt{P_{S_{e_I}}^i P_{S_{e_I}}^i}\sqrt{P_{S_{e_J}}^j P_{S_{e_J}}^j }} \right] 
\end{equation}
and $V^{-1}_v$ is the inverse of the regularized volume operator \cite{Ashtekar:1997fb} $V_v$ at $v$:
\begin{align}
	V_v &\coloneq {\abs{\frac{(\kappa \gamma)^3}{8 \cdot 3!} \sum \limits_{I,J,K} \det(\vec{e}_I,\vec{e}_J,\vec{e}_K) \epsilon_{ijk} P_{S_{e_I}}^i P_{S_{e_J}}^j  P_{S_{e_K}}^k}}^{1/2}
\end{align}
such that $\vec{e}$ denotes the normalized tangent vector to the edge $e$ at $v$.
We underline that the Lorentzian part $C_{L}$ is only a function of the fluxes $P$'s (around the vertex $v$) and does not depend on the holonomies $h$'s.
Thus, at the quantum level, while the Euclidean part $C_{E}$ will induce spin shifts when acting on spin network states, the Lorentzian part $C_{L}$ will not generate spin shifts. In this sense, $C_{E}$ may be interpreted as a kinetic term, while $C_{L}$  plays the role of a potential depending on those spins.

This concludes the regularization procedure of the classical Hamiltonian. The regularized expression can then be promoted to quantum operators acting on the spin network states, and hence completing the implementation of the canonical dynamics of general relativity in the quantum theory. In the next parts, we will focus on the description of the phase space associated to a fixed graph and the corresponding LQG induced dynamics. This will lay the ground for section \ref{CandyModel} where we study this dynamics within the classical candy graph model.

%%%%%%%%
\subsection{Phase space of a fixed graph and the spinor parametrization of LQG}\label{I.B}
%%%%%%%%

As noted earlier, for each graph $\Gamma$ with $n_e$ edges, the Hilbert space ${\cal H}_\Gamma$ corresponds to $L^2 (SU(2)^{n_e},d\mu_H)$. Independently of the construction presented above, the space $L^2 (SU(2)^{n_e},d\mu_H)$ can be viewed as the quantization of a classical phase space defined as $[T^* SU(2)]^{n_e}$, the cotangent bundle over the group $SU(2)$, which in turn is isomorphic to $[SU(2) \times \su(2)]^{n_e}$. Therefore, the truncation of the LQG Hilbert space to a graph $\Gamma$ can be mapped to a classical formulation with $[SU(2) \times \su(2)]^{n_e}$ as a phase space, and a Hamiltonian given by the regularized expressions above \eqref{LQG.Ham} and \eqref{LQG.Ham.Parts}. This classical system can then be studied in its own right, and its dynamical properties can provide semi-classical interpretations to its quantum counterpart. In the following, we present the spinor formalism used to parametrize the $[SU(2) \times \su(2)]^{n_e}$ phase space and identify the LQG induced Hamiltonian governing the dynamics.

Given a graph $\Gamma$ with $n_e$ edges, the corresponding phase space $[T^* \SU(2)]^{n_e} \simeq  [\SU(2) \times \su(2)]^{n_e}$ can be parametrized in terms of $n_e$ pairs of variables $(g_e, \bX_e) \in \SU(2)\times \su(2)$, each associated to an edge $e \in \Gamma$. 
We decompose the Lie algebra elements as $\bX_e = -i X_e^a \sigma_a$ ($a = 1, 2, 3$) with $X_e^a \in \mathbbm{R}$. The Pauli matrices $\sigma_a$ are normalized such that $\sigma^{a}\sigma^{b} = \delta^{ab} \mathbbm{1} + i \epsilon^{abc} \sigma^c$.
The phase space is endowed with the following symplectic structure
\begin{align}\label{Symp.gX}
	\{g_k,g_l\} = 0 \,,\qquad \{X_k^{a},X_l^{b}\} = \delta_{kl} \epsilon^{abc}X_l^{c} \,,\qquad \{X_k^{a},g_l \} = -\frac{i}{2} \delta_{kl} g_l \sigma^{a}
\end{align}

There is another parametrization of this phase space, the {\it spinor parametrization} \cite{Livine:2011gp, Livine:2011zz}, which often turns out to be more convenient in performing calculations and better suited for the interpretation of the spin network states in terms of simplicial geometries. The spinor parametrization consists of associating a pair of spinors $\{\ket{z}, \ket{w}\}$ to the initial and the final vertices of each edge of the graph respectively. 

Using the standard symplectic structure on $\mathbbm{C}^2 \times \mathbbm{C}^2$, and by requiring that two spinors on opposite sides of the same edge have equal norms, one can reconstruct the variables $(g, X) \in SU(2)\times \su(2)$ for the edge with the symplectic structure \eqref{Symp.gX}, hence recovering the loop gravity phase space for one edge \cite{Livine:2011gp}.

In the spinor formalism for LQG, we define a spinor $\ket{z} \in \mathbbm{C}^2$, its conjugate $\bra{z}\in \mathbbm{C}^2$ and the spinor $\coket{z} \in \mathbbm{C}^2$ as:
\begin{align}\label{Spinors}
	\ket{z}=\mat{c}{z^0\\z^1}, \qquad
	\bra{z}=\mat{cc}{\bar{z}^0 &\bar{z}^1},\qquad \coket{z}\coloneq-i\sigma^2\ket{\bar{z}} = \mat{c}{- \bar{z}^1\\ \bar{z}^0}, \quad \text{with}\quad\sigma^2=\mat{cc}{0 & -i\\ i & 0}.
\end{align}
Each spinor variable is provided with the canonical Poisson bracket
\begin{align}
	\{z^{A},z^{B}\}=\{\bz^{A},\bz^{B}\}=0,\qquad \{z^{A},\bz^{B}\}=-i\delta^{AB}\,, \quad
	\text{with}\quad A, B=0,1\,;
\end{align}
and it defines a 3d real vector by its projection onto the Pauli matrices :
\begin{align}
	X^{a}=\frac{1}{2}\bra{z} \sigma^{a}\ket{z}=\frac{1}{2}\sigma^{a}_{AB}\bz^{A} z^{B},\qquad
	X \equiv \frac{1}{2}\braket{z|z}\,.
\end{align}
Notice that the spinor $\ket{z}$ is entirely determined by the corresponding 3-vector $\vec X$, up to a $U(1)$ factor. This phase has been shown to play an important role in the geometric interpretation of the space of intertwiners as wave-functions over the space of a classical polyhedron \cite{Freidel:2010tt}. As we discuss later in section \ref{Beyond2leg}, the phase can be used to introduce an extension of the Candy graph model which we will be studying in sections \ref{CandyModel} and \ref{Solutions}.

The vector components $X^a$ naturally form a $\su(2)$ Lie algebra,
\begin{align}
	\{X,X^{a}\}=0\,,\qquad
	\{X^{a},X^{b}\} = \epsilon^{abc}X^{c}\,,
\end{align}
whose Casimir is the squared norm $\vec{X}^{2}=X^{2}$, and which generates the $SU(2)$ group action on the spinors
\begin{align}
	e^{{\{\theta \hat{u}\cdot\vec{X}, \bullet\}}}\,\ket{z} = g(\theta,\hat{u})\,\,\ket{z}
	\,,
	\qquad
	g(\theta,\hat{u}) = e^{i\frac\theta2\,\hat{u}\cdot\vec{\sigma}} \in SU(2)\,.
\end{align}
Now, given two spinors $\ket{z}$ and $\ket{w}$ associated to an edge linking two vertices, we define a $SU(2)$ group element $g$ mapping the spinors from one vertex to the other\footnotemark:
\begin{gather}
	g=\frac{\coket{w}\bra{z}-\ket{w} \cobra{z}}{\sqrt{\braket{z|z} \braket{w|w}}}
	\,, \label{g}\\[2ex]
	g\,\frac{\ket{z}}{\sqrt{\braket{z|z}}}=\frac{\coket{w}}{\sqrt{\braket{w|w}}}
	\,,\quad
	g\,\frac{\coket{z}}{\sqrt{\braket{z|z}}}=\,-\frac{\ket{w}}{\sqrt{\braket{w|w}}}
	\,.
\end{gather}
\footnotetext{One can also define ``boosted'' holonomies as
\be
g^{(\lambda)}
=
\f{e^{+\lambda}\,|w]\la z|\,-\,e^{-\lambda}\,|w\ra [z|}{\sqrt{\la w|w\ra\la z|z\ra}}\in\SL(2,\C)
\,,\nn
\ee
obtained by an inverse dilatation on the two spinors, $|z\ra\mapsto e^{-\f\lambda 2}|z\ra$ and $|w\ra\mapsto e^{+\f\lambda 2}|w\ra$, for $\lambda \in \R$.
Up to the subtlety that $g^{(\lambda)}$ is not in $\SU(2)$ but more generally in $\SL(2,\C)$, it satisfies the same Poisson algebra as the original holonomy:
\be
\{g^{(\lambda)},g^{(\lambda)}\}\sim0
\,,\quad
\{X_{a},g^{(\lambda)}\}=-\f i2\,g^{(\lambda)}\sigma_{a}
\,,\quad
\{Y_{a},g^{(\lambda)}\}=+\f i2\,\sigma_{a}g^{(\lambda)}
\,.\nn
\ee
Changing $\lambda$ is thus a canonical transformation, very similar in spirit to the Immirzi-Barbero parameter, generated by canonical transformations on the continuous fields changing the relative weight of the 3d spin-connection and extrinsic curvature in the Ashtekar-Barbero connection. It might be interesting to understand if this is more than a mere similarity.
}
Let us call $X^{a}$ the 3-vectors defined by the spinors $\ket{z}$ and $Y^{a}$ the 3-vectors defined by the spinors $\ket{w}$. Then, if we assume norm-matching condition along each edge,
\begin{align}\label{NormMatching}
	{\cal C}=\braket{z|z}-\braket{w|w}=0\,,
\end{align}
$X^{a}$, $Y^{a}$ and $g$ form a $T^{*} SU(2)$ Lie algebra:
\begin{align}
	\{g,g\} = 0
	\,,\qquad
	\{X^{a},g\} = g\,\left(\frac{-i\sigma^{a}}{2}\right)
	\,,\qquad
	\{Y^{a},g\} = \left(\frac{+i\sigma^{a}}{2}\right)\,g
	\,,\qquad
	\{X^{a},Y^{b}\}=0.
\end{align}
Thus we recover the $SU(2)\times \su(2)$ parametrization of $T^* SU(2)$ with the appropriate symplectic structure.

Note that since we often prefer working on the $SU(2)$-gauge invariant Hilbert space, one must additionally impose the Gauss constraint at each vertex of the graph. This constraint translates into a closure constraint at each vertex in terms of the 3-vectors $X^{a}$, namely
\begin{align}\label{GaussX}
	\mathcal{X}^a \equiv \sum_k X_k^a = 0\,, \qquad \text{for every }a \in \{1,2,3\}
\end{align}
where $k$ runs through the edges meeting at a given vertex.

%%%%%%%%
\subsection{Dynamics on the phase space of a fixed graph}\label{I.C}
%%%%%%%%

As discussed earlier, the Hamiltonian formulation of general relativity encodes the dynamics in constraints. In this case, quantities such as the metric are not observable.  This translates into a frozen picture where there is no time flow nor evolution of physical quantities. This specific aspect raises
several issues in the interpretation of a quantum theory of gravity. This problem can be circumvented in certain cases, for instance when considering a particular solution of Einstein equations such as homogeneous spacetimes, where one chooses a specific coordinate system and quantizes the model in this coordinate system. This is equivalent to fixing the lapse and shift parameters and ensuring that their expressions are preserved by the dynamics. The drawback is of course the fact that one breaks diffeomorphism invariance and the interpretation of the physics derived within the model would be tied to the choice of coordinates system. Furthermore, a particular coordinate system, which corresponds to a chosen set of observers, can only be valid for specific spacetime configurations. This situation can be improved by using matter fields as coordinates, this is the so-called deparametrization approach. Deparametrization requires coupling adequate matter fields to gravity, such that the constraints of the theory, in particular the scalar constraint, can be solved for some of the degrees of freedom of the matter fields. This consequently allows to use matter degrees of freedom as coordinates which parametrize the physical evolution of the remaining degrees of freedom in the system, including the gravitational ones. 
There have been many models of deparametrized gravity \cite{Kuchar:1991pq, Kuchar:1990vy, Rovelli:1993bm, Brown:1994py, Kuchar:1995xn, Giesel:2012rb}, which lead to theories where gravity is fully quantized. These models become a very rich playground to test the many technical steps of the quantization procedures along with the development and analysis of new methods and ideas to answer even more complex questions concerning the semi–classical and continuum limits of the quantum gravity theory.

That being said, the gravitational part of the Hamiltonian constraint $H(N)$ is generally understood as the generator of diffeomorphisms along the time direction. In the terms of gauge theories, the gravitational Hamiltonian generates gauge orbits under the action of infinitesimal time translations. In the models we investigate, which can be viewed as truncations of the full theory, we approach the dynamics from the perspective of trying to understand the Hamiltonian flow generated by the gravitational part and encoded in the Poisson brackets $\{H(N), \cdot \}$. This is because the analysis of the Hamiltonian flow provides a universal understanding of time evolution which goes beyond vacuum gravity, and extends to contexts where matter fields can be included. Therefore, as far as our analysis goes, we are not concerned with solving the Hamiltonian constraint $H(N)=0$, instead, we are looking within all possible trajectories and initial configurations without any fixed value of $H(N)$.

Let us now turn to the question of defining a Hamiltonian on the phase space of a fixed graph. This means that we would like to introduce a Hamiltonian expressed in terms of the spinor variables. To do so, we use the regularized version (pre-quantization) of the Hamiltonian $H(N)$ in \eqref{LQG.Ham} and \eqref{LQG.Ham.Parts}. The expression of the Hamiltonian $H(N)$ is a sum of terms of the form
\begin{align}
	c_1 \epsilon_{ijk} \text{Tr} \left[ g_{IJ} \tau^i \right] P_I^j P_J^k V^{-1}
	+ c_2 \sqrt{ (\epsilon_{ijk} P_I^j P_J^k) (\epsilon_{i j' k'} P_I^{j'} P_J^{k'})} \left( \frac{2\pi}{\lambda_{IJ}} - \Theta_{IJ} \right) V^{-1} \label{Regularized.Hamiltonian}
\end{align}
where $c_1$ and $c_2$ are constants, $g_{IJ}$ is a group element around a loop and $P_I^i = P_{S_{e_I}}^i$ are the fluxes. The first term in \eqref{Regularized.Hamiltonian} is proportional to a trace of a group element around a loop, contracted with the corresponding fluxes, and can be interpreted as a kinetic term. While the second term is of quadratic order in the fluxes and does not involve $SU(2)$ group elements, therefore it can be interpreted as a potential term. Looking back at the construction of the spinor formalism reviewed in the previous section, we can infer the following mappings between the quantities in \eqref{Regularized.Hamiltonian} and the spinor variables, namely
\begin{align}\label{Hamiltonian.mappings}
	&P_I^i \longrightarrow X_{I}^i\ , \quad
	\sqrt{ (\epsilon_{ijk} P_I^j P_J^k) (\epsilon_{i j' k'} P_I^{j'} P_J^{k'})} \longrightarrow \abs{\vX_{I} \wedge \vX_{J}}\ , \quad 
	\epsilon_{ijk} P_I^i P_J^j P_K^k \longrightarrow \vX_{I} \cdot (\vX_{J} \wedge \vX_{K}) \\[2ex]
	&\epsilon_{ijk} P_I^i P_J^j \text{Tr} \left[ g_{IJ} \tau^j \right] \longrightarrow 2i \epsilon_{ijk} X_{I}^i X_{J}^j \tr \left[ \, \sigma^k g_{IJ} \right] = 2
	\tr \left[ ((\vX_{I}\cdot\vsigma)\,(\vX_{J}\cdot\vsigma)-\vX_{I}\cdot\vX_{J})\,g_{IJ} \right]
\end{align}
The above expressions provide a straightforward way to construct an LQG induced Hamiltonian on the spinor phase space associated to a fixed graph, and we obtain that the regularized Hamiltonian in \eqref{Regularized.Hamiltonian} produces a discrete Hamiltonian of the form
\begin{equation}\label{Discrete.Hamiltonian}
	H(N) = \sum \limits_{v \in \Gamma} \frac{N(v)}{V(v)} \left( H_E(v) + H_L(v) \right)
\end{equation}
where
\begin{align}
	V(v) &= \left[\frac{1}{3!} \sum \limits_{I\neq J\neq K} \abs{\vX_{I} \cdot (\vX_{J} \wedge \vX_{K})} \right]^{1/2} \\[2ex]
	H_E(v) &= c_1(v) \sum \limits_{I\neq J} 2 \tr \left[ ((\vX_{I}\cdot\vsigma)\,(\vX_{J}\cdot\vsigma)-\vX_{I}\cdot\vX_{J})\,g_{IJ} \right] \label{HE}\\[2ex]
	H_L(v) &= c_2(v) \sum \limits_{I\neq J} \abs{\vX_{I} \wedge \vX_{J}} \left( \frac{2\pi}{\lambda_{IJ}} -  \pi - \arccos \left[ \frac{\vX_{I} \cdot \vX_{J}}{\abs{X_I}\abs{X_J}} \right]  \right) \label{HL}
\end{align}

As expected, the Hamiltonian in \eqref{Discrete.Hamiltonian} is not polynomial in the $X$ variables, in particular the part $H_L$, which is a direct consequence of the fact that the original continuous Hamiltonian in \eqref{AB.Const.} is not polynomial in $E$. Naturally, the equations of motion generated by this Hamiltonian for a generic observable would be complicated, and although they may often be solvable numerically, analytical solutions are elusive. As mentioned above, the term $H_L$ can be interpreted as a potential term controlled by the parameters $c_2$, while the term $H_{E}$ can be interpreted as playing the role of a kinetic term. In fact, when the parameters $c_1$ and $c_2$ are chosen adequately\footnote{The parameters $c_1$ and $c_2$ are induced by the averaging coefficients $k_E(v)$ and $k_E(v)$ in the regularized expressions in \eqref{LQG.Ham.Parts}. Each of these coefficients is actually defined only up to an undetermined constant factor, which is often set to one, but it does not have to be. This is one of the ambiguities which generically arise in the quantization of classical functionals in LQG, see for instance \cite{Ashtekar:1997fb} for the simpler case of the volume. Ideally, such ambiguity should be taken care of in the context of a renormalization scheme, where these parameters would run. However, since this is beyond the scope of the present work, it is not unreasonable to anticipate using these parameters to study the Lorentzian dynamics in a perturbative fashion. All with the objective of gaining concrete insights about the dynamics of the full theory.}, one can cast $H_L$ as a time-independent perturbation of the unperturbed Hamiltonian defined as $H_E$, and consequently setting $H_E$ as the Hamiltonian of an idealized system which needs to be studied first, before moving to the perturbed case.

Furthermore, our analysis of the candy graph, presented in the next sections, will focus on the evolution of areas, given by the norm of the 3d flux vectors $\vec X$. Since those norms commute with the vector components, they will automatically commute with the Hamiltonian term $H_L$, as well as with the volume $V$, that is $\{X, H_L\} = \{X, V\} =0$. Thus we can safely put aside both $H_{L}$ and $V$ terms, for now, as long as we focus on the dynamics of the areas (which become the spins at the quantum level). We will therefore only consider the term $H_{E}$ in our analysis of the dynamics.
Of course, the terms $H_{L}$, $V$ and other possible geometrical terms, will be become (highly) relevant when investigating the dynamics of angular variables and the direction of the flux vectors.
We will discuss going beyond this simplification later in section \ref{Beyond2leg}.

Additionally, we choose the lapse function $N$ such that $N(v) \propto V(v)$. This choice of conformal gauge actually simplifies the Hamiltonian, as the lapse then cancels the inverse volume factor in the expression of $H$. This  lapse corresponds to a specific choice of ``time coordinate" which parametrizes the orbits generated by the Hamiltonian.
Since we are primarily interested in the properties of the flow generated on the candy graph, and that there is no preferred notion of proper time at this stage, nothing is lost by carrying out our analysis in a specific gauge.

In summary, the term $H_E$ will be the starting point for the Hamiltonian that we consider in the context of the candy graph model, and we will be able to solve the equations of motion analytically for the area observables we are interested in.

This concludes the review of the spinor formulation of the phase space of a fixed graph and the associated dynamics. In the next section, we introduce the candy graph model \cite{Aranguren:2022nzn}. This model is defined by the phase space associated to a graph with two vertices, with only two edges connecting them. After introducing a Hamiltonian based on the term $H_E$ in \eqref{HE}, we study the induced equations of motion for the spinor variables and determine the general solutions and their properties. This analysis allows to identify different regimes of the evolution dictated by the LQG dynamics.

%%%%%%%%
\section{The Candy Graph model}\label{CandyModel}
%%%%%%%%

\begin{figure}[!ht]
    \centering
    \includegraphics[height=25mm]{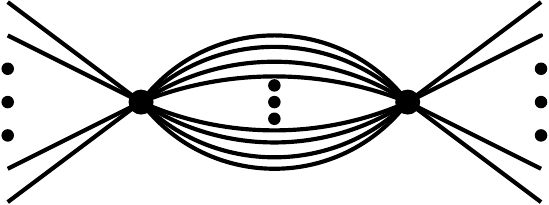}
    \caption{\small The {\it Candy Graph} as the simplest building block of spin network states.}
    \label{fig:candy}
\end{figure}

Let us consider the simplest block of a spin network structure, consisting of 2 nodes linked by possibly several spin network edges and with an arbitrary number of external edges. This is the most basic piece of a spin network. It is a generalization of the 2-vertex model previously studied in \cite{Borja:2010gn,Livine:2011up,Aranguren:2022nzn,Cendal:2024uzu} to an open graph including a boundary. We call it "candy graph" as illustrated on fig.\ref{fig:candy}.
This simple configuration allows for local curvature excitations developing between the two nodes. It becomes possible to study analytically and numerically the local dynamics of the curvature in LQG. We would propose to consider it as the equivalent of the harmonic oscillator for the theory, more precisely as the proof-of-concept of the LQG dynamics.

More to the point, since this basic graph has a boundary and open links, it opens the door to the controlled study of dynamics in terms of the boundary state.
At the classical level, it would allow to investigate the various choices of LQG Hamiltonians, but also the various choices of boundary data -- fixed boundary data, given time-dependent boundary data, dynamically-evolving boundary data (with possible incoming and outgoing symplectic flux) -- and how they affect the bulk dynamics of the curvature building between the two nodes.
At the quantum level,  it would enable to proceed to a controlled study of the quantization, of the regularization of the classical Hamiltonian constraints and the classification of possible quantum Hamiltonians according to their spectrum, and hopefully also to the in-depth analysis of the quantum counterpart of the various choices of classical boundary data.
Then the open links of the candy graph would allow to glue candy graphs together, possibly opening the door to more complex, potentially scalable, spin network architectures.

Finally, since we can look directly into the dynamics of basic local curvature excitations and study the scale, amplitude and relevance of geometrical fluctuations in a bounded region of space, this would provide original insights into the renormalization flow of Loop Quantum Gravity.
More precisely, for a fixed boundary state, and specifically fixed boundary areas on the open links,  we envision two scenarios for the classical evolution:
\begin{enumerate}
	
	\item we have bounded volume trajectories, most likely oscillatory, which would translate into bounded states at the quantum level with an energy spectrum, similarly to elementary atoms;
	
	\item we have unbounded evolution for the bulk volume and bulk areas, similar to scattering states in particle physics or non-compact cosmological evolution, signaling divergent UV fluctuations which do not affect the exterior dynamics, thus to be renormalized.
	
\end{enumerate}
As often in research into quantum gravity, the dynamics could, in the end, be less binary and reveal a richer phenomenology, allowing for further insights into the coarse-graining flow and renormalization of Loop Quantum Gravity.

%%%%%%%%
\subsection{The one-loop phase space}
%%%%%%%%

\begin{figure}[!ht]
    \centering
    \begin{subfigure}[b]{\textwidth}
    		\centering
		\includegraphics[width=120mm]{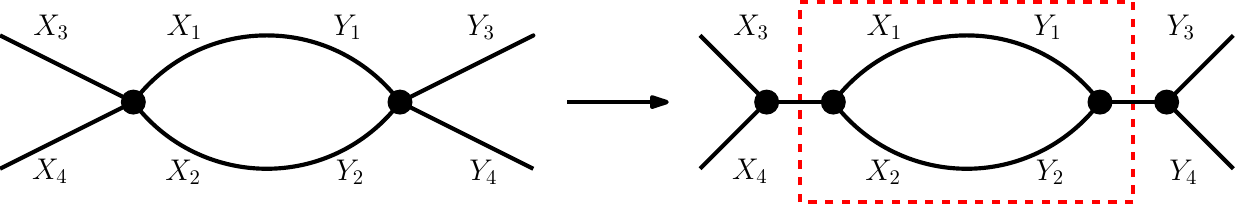}
        \caption{\small Unfolding and stretching the 4-valent nodes.}
    \end{subfigure}
    \begin{subfigure}[b]{\textwidth}
    	\centering
		\includegraphics[width=55mm]{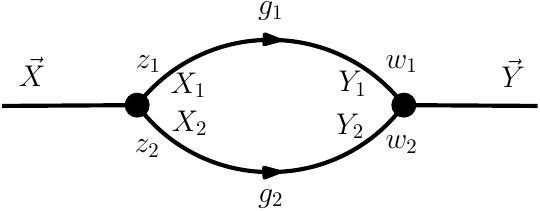}
        \caption{\small Simplified candy with 2 external legs}
    \end{subfigure}
    \caption{\small Candy graph with two 4-valent nodes: to focus on the dynamics of the bulk loop, and thus of the local curvature excitation,  we freeze the intertwiner degrees of freedom by stretching the nodes and keeping the norms $|\vX|$ and $|\vY|$ fixed.}
    \label{fig:candysettting}
\end{figure}

We consider a simplified Candy graph system, where the two vertices $v_X$ and $v_Y$ are connected by two edges, and we study the dynamics of the bulk only. This allows to reduce the problem of the external links to a single internal link per node, using the fact that it is enough to have the vector representing the sum of the vectors associated to the external link. Schematically, this can viewed as stretching the nodes as drawn on fig.\ref{fig:candysettting}. This setup is perfectly suitable if we put aside the dynamics of the intertwiners, for instance by fixing the boundary state (i.e.~the states of the boundary spins). We call this model the 2-leg Candy graph model.
Focusing on the bulk degrees of freedom, we denote the spinors associated to the two edges as $\ket{z_i}$ at the vertex $v_x$ and $\ket{w_i}$ at the second vertex $v_y$, $i \in \{1,2\}$, and we have
\be
\vX_{1}=\f12\la z_{1}|\vsigma|z_{1}\ra
\,,\quad
\vX_{2}=\f12\la z_{2}|\vsigma|z_{2}\ra
\,,\quad
|z_{a}\ra\la z_{a}|=X_{a}\id+\vX_{a}\cdot\vsigma\,,
\ee
\be
\vY_{1}=\f12\la w_{1}|\vsigma|w_{1}\ra
\,,\quad
\vY_{2}=\f12\la w_{2}|\vsigma|w_{2}\ra
\,,\quad
|w_{a}\ra\la w_{a}|=Y_{a}\id+\vY_{a}\cdot\vsigma\,,
\ee
The group elements representing the holonomies along the edges are given by \eqref{g} as
\be
g_{1}
=
\f{|w_{1}]\la z_{1}|-|w_{1}\ra [ z_{1}|}
{\sqrt{\la z_{1}|z_{1}\ra\la w_{1}|w_{1}\ra}}
\,,\qquad
g_{2}
=
\f{|w_{2}]\la z_{2}|-|w_{2}\ra [ z_{2}|}
{\sqrt{\la z_{2}|z_{2}\ra\la w_{2}|w_{2}\ra}}
\,.
\ee
The norm-matching contraints \eqref{NormMatching} imply
\be
\la z_{1}|z_{1}\ra=\la w_{1}|w_{1}\ra
\,,\qquad
\la z_{2}|z_{2}\ra=\la w_{2}|w_{2}\ra
\,.
\ee
The Gauss constraints with open links at both nodes, imposing to $\SU(2)$-invariance or more precisely covariance under $\SU(2)$-invariance, imply
\be
\vX_{1}+\vX_{2}=\vX
\,,\qquad
\vY_{1}+\vY_{2}=\vY
\,.
\ee
where $\vec X$ and $\vec Y$ are the vectors associated to the internal links at the two vertices as depicted in fig.\ref{fig:candysettting}. This completes the description of the kinematics of the 2-leg candy graph model, and we can move to the implementation of the dynamics.

%%%%%%%%
\subsection{The bulk Hamiltonian in the 2-leg Candy graph model}
%%%%%%%%

As discussed at the end of section \ref{I.C}, we can consider the Euclidean term $H_E$ in \eqref{HE} to be the Hamiltonian which generates the dynamics for the bulk degrees of freedom in the 2-leg Candy graph model. To simplify the analysis, we introduce the following functions
\begin{align}
E &= \la z_{1}|z_{2}\ra \la w_{1}|w_{2}\ra ,\qquad \bE = \la z_{2}|z_{1}\ra \la w_{2}|w_{1}\ra \\
F &= [ z_{1}|z_{2}\ra [ w_{1}|w_{2}\ra ,\qquad \bF = \la z_{2}|z_{1}] \la w_{2}|w_{1}] 
\end{align}
We can then express $H_E$ in terms of the above functions by calculating the two traces in \eqref{HE}. For the vertex $v_x$ we obtain
\begin{align}
	\tr \left[ (\vX_{1}\cdot\vsigma)\,(\vX_{2}\cdot\vsigma)\,g_{2}^{-1}g_{1} \right] &= \f{1}{4}[(E+\bE)-(F+\bF)] \\[2ex]
	(\vec X_{1} \cdot \vec X_{2})\ \tr\left[g_{2}^{-1}g_{1}\right] &= \frac{\vec X_{1} \cdot \vec X_{2}}{4X_{1}X_{2}} \big{[} (E+\bE)+(F+\bF) \big{]}
\end{align}
therefore
\begin{align}
	H_E(v_x) \propto \left[ 1 - \frac{\vec X_{1} \cdot \vec X_{2}}{4X_{1}X_{2}} \right] (E+\bE) - \left[ 1 + \frac{\vec X_{1} \cdot \vec X_{2}}{4X_{1}X_{2}} \right] (F+\bF) \label{HE_CG}
\end{align}
The same expression holds for the vertex $v_y$ by simply replacing the variables $\vec X_i$ by $\vec Y_i$.

\medskip

Now, the full Hamiltonian $\mathbf{H}=H_{E}+H_{L}$ would involve both Euclidean and Lorentzian terms. However,  as we discussed earlier, the Lorentzian term $H_{L}$ only depends on the flux vectors and does not involve any holonomy. It thus commutes with the vector norms and is not relevant
to the  evolution of the areas $X_1 = Y_1$ and $X_2 = Y_2$ associated to the two bulk edges.
Moreover, we will slightly simplify the Euclidean term and consider
a simplified form of the Hamiltonian, namely
\be
\label{Hamiltonian.Ansatz}
H = \alpha E+\balpha\bE +\beta F+\bbeta \bF
\,,\qquad
\textrm{with} \quad \alpha, \beta \in \mathbbm{C}\,,
\ee
where we have replaced the factors depending on the vectors $\vec X$ by constant complex parameters.

As much as $H_E$, the Hamiltonian in \eqref{Hamiltonian.Ansatz} is invariant under phase transformation along edges and under $\SU(2)$ transformations at nodes. It can be viewed as a straightforward generalization of dynamics of 2-vertex graphs studied in \cite{Borja:2010gn,Borja:2010rc,Livine:2011up, Aranguren:2022nzn,Cendal:2024uzu,Garay:2025bqk,Garay:2025cis}, but allowing for sources in the Gauss laws, i.e.~$\vX_{1}+\vX_{2}=\vX\ne 0$, which can be interpreted as torsion defects or tags in the terminology of \cite{Charles:2016xwc}.

It also follows naturally from two paths:
\begin{itemize}
\item if one argues that, in loop quantum gravity, the holonomy-flux variables, and thus the spins and spinors, are the relevant degrees of freedom to describe the geometry and its evolution, then it is natural to expect the Hamiltonian to be polynomial in those variables. The Hamiltonian $H$ is the most general such polynomial ansatz  at lowest order compatible with the $\SU(2)$ gauge invariance requirement (that is quartic in the spinor variables).

\item This Hamiltonian ansatz is also a straightforward combination of $\Tr XX g_{2}^{-1}g_{1}$ terms. It is thus a slight variation around $H_{E}$, which might be interpreted as a different discretization/regularization of the continuum Hamiltonian.
To realize this explicitly, let us first introduce shifted holonomy variables,
\be
g^{(\theta)}
=
\f{e^{-i\theta}\,|w]\la z|\,-\,e^{+i\theta}\,|w\ra [z|}{\sqrt{\la w|w\ra\la z|z\ra}}\in\SU(2)
\,,
\ee
obtained by an inverse phase transformation on the two spinors. This is a mere canonical transformation and this shifted holonomy satisfies the same Poisson algebra with the flux vectors as the original holonomy.
Then, we compute:
\be
H=\Tr\bigg{[}\big{(}
\lambda X_{1}X_{2} +\mu (\vX_{1}\cdot\vsigma)(\vX_{2}\cdot\vsigma) +2i\rho (X_{1}((\vX_{2}\cdot\vsigma)+(\vX_{1}\cdot\vsigma)X_{2})
\big{)}\,(g_{2}^{(\theta)})^{-1}g_{1}^{(\theta)}
\bigg{]}
\ee
with the identification between constants given by $4\alpha=\lambda+\mu+i\rho$ and $4\beta=(\lambda-\mu)e^{+2i\theta}$.
\end{itemize}

Now that we have a suitable and well defined Hamiltonian, let us turn to the derivation of the equations of motion for the areas and their solutions.

%%%%%%%%
\subsection{Area evolution as a Non-Linear Schrodinger equation}
%%%%%%%%

We are interested in determining the evolution of the bulk areas $X_{1}=\f12\la z_{1}|z_{1}\ra$ and $X_{2}=\f12\la z_{2}|z_{2}\ra$, keeping in mind the norm-matching conditions along the two bulk links, $X_{1}=Y_{1}$ and $X_{2}=Y_{2}$. We then compute the Poisson brackets of the two bulk areas with the terms of the Hamiltonian \eqref{Hamiltonian.Ansatz}, obtaining:
\be
\{X_{1},E\}=-\f i2E
\,,\quad
\{X_{1},\bE\}=+\f i2\bE
\,,\quad
\{X_{2},E\}=+\f i2E
\,,\quad
\{X_{2},\bE\}=-\f i2\bE
\,,
\ee
\be
\{X_{1},F\}=+\f i2F
\,,\quad
\{X_{1},\bF\}=-\f i2\bF
\,,\quad
\{X_{2},F\}=+\f i2F
\,,\quad
\{X_{2},\bF\}=-\f i2\bF
\,.
\ee
The action of each term of the Hamiltonian \eqref{Hamiltonian.Ansatz} is clearer when recast in terms of the total bulk area $A\equiv X_{1}+X_{2}$ and the area difference $a\equiv X_{1}-X_{2}$. This is the same as separating the 2-body trajectory in terms of the center of mass motion and the relative motion, and we get
\be
\begin{array}{llll}
\{A,E\}=0
\,,\quad&
\{A,\bE\}=0
\,,\quad&
\{A,F\}=+iF
\,,\quad&
\{A,\bF\}=- i\bF
\,,\\
\{a,E\}=-iE
\,,\quad&
\{a,\bE\}=+i\bE
\,,\quad&
\{a,F\}=0
\,,\quad&
\{a,\bF\}=0
\,,
\end{array}
\ee
This provides a clear identification of the role of the $E$ and $F$ parts of the Hamiltonian:
on the one hand, the $E$-terms do not affect the total area but create motion of the difference in area excitations  between the two links. Since the areas are bounded from below, the motion inevitably consists of oscillations in the area difference $a$.
On the other hand, the $F$-terms do not change the area difference, but drive the evolution of the total area.
This is consistent with the previous works  \cite{Borja:2010gn,Borja:2010rc,Livine:2011up} on the 2-vertex model and we obtain the following equations of motion:
\be
\dot{A}
=\{A,H\}=\{A,\beta F+\bbeta \bF\}
=i(\beta F-\bbeta \bF)
\,,
\ee
\be
\dot{a}
=\{a,H\}=\{a,\alpha E+\balpha\bE \}
=i(-\alpha E+\balpha\bE )
\,,
\ee
where ``$\ \dot{}\ $" denote the derivative with respect to the time parameter $t$ along the trajectories generated by the Hamiltonian $H$. We observe that in order to obtain closed equations of motion, we need to take a second time derivative and thus another Poisson bracket with the Hamiltonian. This requires computing the brackets between the four terms of the Hamiltonian, and it turns out that these can be simplified using the area-matching constraints and the Gauss laws with external edges.

The first striking result is that the $E$ and $F$ terms commute in our setting (with only two bulk links):
\beq
\{E,F\}
&=&
\{\la z_{1}|z_{2}\ra \la w_{1}|w_{2}\ra ,[z_{1}|z_{2}\ra [ w_{1}|w_{2}\ra \}
\nn\\
&=&
i\left(
[z_{2}|z_{2}\ra \la w_{1}|w_{2}\ra[ w_{1}|w_{2}\ra
+
[w_{2}|w_{2}\ra\la z_{1}|z_{2}\ra [z_{1}|z_{2}\ra
\right)\\
&=&0\,\nn
\eeq
and more generally
\be
\{E,F\}=\{E,\bF\}=\{\bE,F\}=\{\bE,\bF\}=0
\,.
\ee
This implies that the evolution of the areas difference $a$ fully decouples from the evolution of the total area $A$.
The remaining non-vanishing brackets are given as
\beq
\{E,\bE\}
&=&
-i\left(\la z_{1}|z_{1}\ra-\la z_{2}|z_{2}\ra \right) \la w_{1}|w_{2}\ra\la w_{2}|w_{1}\ra
-i\left(\la w_{1}|w_{1}\ra-\la w_{2}|w_{2}\ra \right) \la z_{1}|z_{2}\ra\la z_{2}|z_{1}\ra
\nn\\[1ex]
&\underset{X_{a}=Y_{a}}\sim&
-i\,2a\left(\la w_{1}|w_{2}\ra\la w_{2}|w_{1}\ra+\la z_{1}|z_{2}\ra\la z_{2}|z_{1}\ra\right)
\nn\\[1ex]
&\sim&
-4ia(X_{1}X_{2}+\vX_{1}\cdot\vX_{2}+Y_{1}Y_{2}+\vY_{1}\cdot\vY_{2})
\sim
-4ia(B^{2}-a^{2})
\eeq
with $B^{2}\equiv (X^{2}+Y^{2})/2$.
Similarly, we have:
\be
\{F,\bF\}
\sim
-4iA(A^{2}-B^{2})
\,,
\ee
where the $\sim$ means the equality upon assuming the area-matching and Gauss law constraints.
This gives the following 2nd order equations of motion for the bulk areas:
\beq
\ddot{A}
=
8 |\beta|^{2}A (A^{2}-B^{2})
\,, \label{A_eom}
\eeq
\beq
\ddot{a}
=
-8 |\alpha|^{2}a (B^{2}-a^{2})
\,. \label{a_eom}
\eeq
Notice the very elegant feature that both the total area $A$ and the area difference $a$ satisfy the exact same differential equation apart from the frequency factors $|\alpha|^{2}$ and $|\beta|^{2}$  coming directly from the Hamiltonian couplings. We nevertheless wrote them slightly differently to underline the natural inequalities arising from their origin as norm of vectors (and their geometrical interpretation as areas):
\be
\left.
\begin{array}{lcl}
A^{2}=\left(|\vX_{1}|+|\vX_{2}|\right)^{2}
&>& |\vX_{1}+\vX_{2}|^{2}=X^{2}
\\
A^{2}=\left(|\vY_{1}|+|\vY_{2}|\right)^{2}
&>&
|\vY_{1}+\vY_{2}|^{2}=Y^{2}
\end{array}\right\}
\qquad\Longrightarrow\quad
A^{2}>(X^{2}+Y^{2})/2=B^{2}
\,,
\ee
\be
\left.
\begin{array}{lcl}
a^{2}=\left(|\vX_{1}|-|\vX_{2}|\right)^{2}
&<& |\vX_{1}+\vX_{2}|^{2}=X^{2}
\\
a^{2}=\left(|\vY_{1}|-|\vY_{2}|\right)^{2}
&<&
|\vY_{1}+\vY_{2}|^{2}=Y^{2}
\end{array}\right\}
\qquad\Longrightarrow\quad
a^{2}<(X^{2}+Y^{2})/2=B^{2}
\,,
\ee

We then recognize the famous non-linear Schr\"odinger equation. Indeed, writing the one-dimensional Schr\"odinger equation for a wave-function with $|\psi|^{4}$ potential in the Hamiltonian, one obtains
\be
i\hbar \pp_{t}\psi=-\f{\hbar^{2}}{2m}\pp_{x}^{2}\psi+\mu |\psi|^{2}\psi
\ee
Assuming that the wave-function has a given frequency, $\psi=\Psi\,e^{-i\om t}$, and that its amplituude $\Psi$ is real, we get:
\be
\f{\hbar^{2}}{2m}\pp_{x}^{2}\Psi
=
\hbar\om \Psi+\mu \Psi^{3}
\,,
\ee
One can then see that the form of the one-dimensional non-linear Schr\"odinger equation (NLSE) coincides with the evolution equations \eqref{A_eom} and \eqref{a_eom}, where the time coordinate is mapped to the spatial coordinate in the NLSE in addition to a direct mapping between coupling constants.
Thus, one can import all the physics of the NLSE, in particular the solitonic solutions, to solve the LQG dynamics of the 2-leg Candy graph model we consider here.

Before moving to the solving of the equations of motion for the areas difference $a$ and the total area $A$, we conclude this section by discussing the dynamics of the bulk curvature.

%%%%%%%%
\subsection{Curvature Dynamics}
%%%%%%%%

The local curvature is often a very adequate observable to characterize the geometric configurations of the system under consideration.

In the context of the 2-leg Candy graph model, one can quantify the curvature in the bulk by the holonomy around the single bulk loop, more precisely by its trace $\tr\,g_{2}^{-1}g_{1}$, which we can express as:
\be
\tr \,g_{2}^{-1}g_{1}
\,=\,
\f1{4X_{1}X_{2}}
\big{[}
(E+\bE)
+(F+\bF)
\big{]}\,.
\ee
Putting aside the factor $X_{1}X_{2}=(A+a)(A-a)/4$, we focus on the {\it loop holonomy observable} defined as $G\equiv(E+\bE)
+(F+\bF)$. Its evolution is given:
\beq
\dot{G}
&=&
-4i(\balpha-\alpha)a(B^{2}-a^{2})
-4i(\bbeta-\beta)A(A^{2}-B^{2})
\,.
\eeq
The evolution of the holonomy observable is thus obtained as a straightforward integral over time of bulk areas.

For instance, for with real couplings $\alpha,\beta\in\R$, the standard holonomy observable  is a conserved charge along classical trajectories. In that case, the trace of the holonomy $\tr\,g_{2}^{-1}g_{1}$ is simply inversely proportional to the product of bulk areas $X_{1}X_{2}\propto A^{2}-a^{2}$, and it goes to 0 if the total area diverges, $A\rightarrow+\infty$. Keep in mind that this does not mean that the curvature goes to 0, it actually means the exact opposite: it is a maximal curvature, with the holonomy around the loop $g_{2}^{-1}g_{1}$ as far as mathematically possible from the trivial holonomy $\id$ in $\SU(2)$. This is consistent with the geometrical interpretation that if bulk areas grow without control while the boundary area remains fixed, it inevitably signals an unbounded bulk curvature.

We now move to the solving of the equations of motion for the areas difference $a$ and the total area $A$, and the analysis of their solutions.

%%%%%%%%
\section{Oscillations and cosmological-like expansion on the Candy Graph}\label{Solutions}

Let us recall the evolution equations, derived earlier, for the total area $A=X_{1}+X_{2}$ and the area difference $a=X_{1}-X_{2}$:
\be
\ddot{a}
=
-8 |\alpha|^{2}a (B^{2}-a^{2})
\,,
\qquad
\ddot{A}
=
8 |\beta|^{2}A (A^{2}-B^{2})
\,.
\ee
These are non-linear differential equations, but importantly they are decoupled. Here the Hamiltonian coupling constants $\alpha$ and $\beta$, as well as the boundary area data $B^{2}=(X^{2}+Y^{2})/2$, are fixed and they are assumed not to evolve in time.

These two equations are identical (up to the numerical pre-factors $|\alpha|^{2}$ and $|\beta|^{2}$), yet, the resulting behaviors of $A$ and $a$ are very different.
This is due to the difference in the allowed range for those two variables, which is directly reflected in different choices of initial conditions for the trajectories $A(t)$ and $a(t)$.

Indeed, on one hand, the area difference $a$ can take positive and negative real values, bounded in absolute value $|a|\le
\textrm{min}(X,Y)$. In particular, it always remains smaller than $B$, so that $(B^{2}-a^{2})\ge 0$. Then the evolution equation is similar to that of a harmonic oscillator: we have a non-linear oscillator, with $(B^{2}-a^{2})$ playing the role of a dynamical frequency. Qualitatively, when $|a|$ grows in absolute value, $(B^{2}-a^{2})$ decreases and the oscillation slows down, which means that we should get flattened oscillations, with transitions between negative and positive plateaus.

On the other hand, the total area $A$ is always positive, and bounded from below, as $A=X_{1}+X_{2}\ge \textrm{max}(X,Y)$. In particular, it remains always larger than $B$ so that $(A^{2}-B^{2})\ge 0$. Then the acceleration $\ddot{A}$ is always positive.
In contrast to the bounded modes in the relative area $a$, we get unbounded hyperbolic trajectories, similar to scattering modes. These trajectories starting with a contracting phase with $\dot{A}<0$, which slows down progressively and turns into an expanding phase $\dot{A}>0$ with growing acceleration.
This behavior is reminiscent of bouncing FLRW cosmologies, especially in light of the mapping between anisotropic 2-vertex model to cosmology  uncovered in the recent works \cite{Cendal:2024uzu,Garay:2025bqk}.
Nevertheless, to push in this direction, we would need to go beyond the two bulk links between the two vertices and consider a fuller candy graph dynamics with an arbitrary larger number of bulk links. We postpone this extension of our analysis to future investigations.

Overall, the picture we get is made of oscillations of the area difference on top of a hyperbolic evolution of the total area, similar to bounded modes evolving on  an expanding background.
Let us now provide exact analytics for those trajectories.

%%%%%%%%
\subsection{Oscillatory modes for the area variation}
%%%%%%%%

We start by analyzing the evolution of the area difference $a$ driven by the equation of motion:
\be
\ddot{a}
=
-8 |\alpha|^{2}a (B^{2}-a^{2})
\,, \label{a_eom1}
\ee
with the two constants $\alpha\in\C$ and $B>0$.
As we pointed out earlier, this non-linear differential  equation is the same as the non-linear Schr\"odinger equation with $|\psi|^{4}$ potential.
Furthermore, we can identify \eqref{a_eom1} with the defining equation for the Jacobi elliptic functions, interpolating between standard trigonometry and hyperbolic trigonometry.
These functions are closely related to Jacobi $\vartheta$-functions, which are the general solutions for quasi-periodic functions in the complex plane, making them instrumental to solve analytically 2d lattice models and represent modular invariance.

Let us then dine into the details of solving \eqref{a_eom1}. We introduce the Jacobi amplitude defined by the integral:
\be
\cI_{k}(\theta)
=
\int_{0}^{\theta} \f{\rd\vphi}{\sqrt{1-k \sin^{2}\vphi}}
=
\int_{0}^{\sin\theta}\f{\rd u}{\sqrt{1-u^{2}}\,\sqrt{1-k u^{2}}}
\,,
\ee
for $\theta\in[0,\f\pi2]$ and $0\le k\le 1$. The Jacobi amplitude allows to define the Jacobi function $\sn_{k}\Theta$ as
\be
\sn_{k}\big{(}\cI_{k}(\theta)\big{)}=\sin\theta
\,,\quad
\sn_{k}\Theta=\sin\big{(}\cI_{k}^{-1}(\Theta)\big{)}
\,,
\ee
with the periodicity conditions,
\be
\sn_{k}(\Theta+2\tau_k)=-\sn_{k}(\Theta)
\,,\quad
\sn_{k}(\Theta+4\tau_k)=+\sn_{k}(\Theta)
\,,
\ee
where $\tau_k=\cI_{k}(\pi/2)$ is the quarter-period which depends on $k$.
The function $\sn$ is a non-linear deformation of the sine function. It reduces to the sine function at $k=0$, and gets streched as $k$ increases to eventually become the hyperbolic tangent in the $k=1$ limit:
\be
\cI_{0}(\theta)=\theta
\,,\quad
\sn_{0}\Theta=\sin\Theta
\,,\qquad
\cI_{1}(\theta)=\int_{0}^{\theta}\f{\rd \vphi}{\cos\vphi}=\textrm{arcth}\sin\theta
\,,\quad
\sn_{1}\Theta=\tanh\Theta
\,.
\ee
It satisfies the differential equation:
\be\label{diff.eq.sn}
\pp_{\Theta}^{2}\sn_{k}\Theta
=
-(1+k)\sn_{k}\Theta+2k\ \sn_{k}^{3}\Theta
\,.
\ee
and its first derivative satisfies
\be
(\pp_{\Theta}\sn_{k}\Theta)^2
=
(1-\sn_{k}^{2}\Theta)\,(1-k\ \sn_{k}^{2}\Theta)
\,,
\ee
We recognize our 2nd order non-linear differential equation in \eqref{diff.eq.sn}. Actually, both $\sn_{k}$ and $1/\sn_{k}$ are solutions to this equation\footnotemark{}. However, while $\sn_{k}$ is clearly bounded (and can vanish), $1/\sn_{k}$ can not reach 0 and escapes to $\infty$. The former will indeed describe the motion of the area difference $a$, while, as we will explain later, the latter gives the hyperbolic trajectory for the total area $A$.
\footnotetext{
More precisely, both functions $f=\sn_k$ and its properly normalized inversion $f=-1/(\sqrt{k}\,\sn_k)$ satisfy both the 1st and 2nd order non-linear equations:
\be
f''=-(1+k)f+2k\ f^{3}
\,,\qquad
f'=\sqrt{(1-f^{2})(1-k\ f^{2})}
\,.
\nn
\ee
}

\begin{figure}[!ht]
    \centering
        \includegraphics[height=50mm]{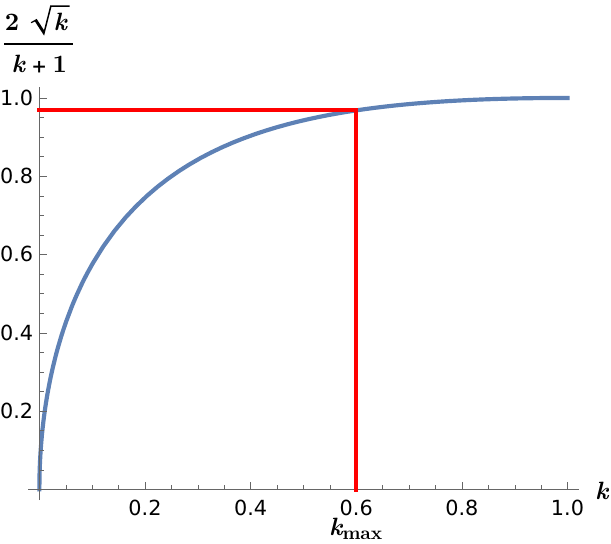}
    \caption{\small Plot of the function $2\sqrt{k}/(1+k)$ showing the upper limit allowed by the boundary data given by the area carried by the external legs of the graph, with maximal value $k_{max}=\textrm{min}(X,Y)/\textrm{max}(X,Y)$. Here, we chose boundary data $X=3$ and $Y=5$, yielding a maximal value of $k_{max}=3/5$. The saturation limit $k_{max}=1$ is  reached when $X=Y$ on the external legs.}
    \label{plot:kfunction}
\end{figure}

Matching the Jacobi elliptic function differential equation with our equation of motion for the area difference \eqref{a_eom1}, we get as classical trajectory:
\be
a(t)=\lambda \,\sn_k (\om t)
\,,\qquad\,\textrm{with}\quad
\om
=
2|\alpha|B\sqrt{\f2{1+k}}
\quad\textrm{and}\quad
\lambda
=
\f{\om \sqrt{k}}{2|\alpha|}
=
B\sqrt{\f{2k}{1+k}}
\,.
\ee
This solution is generated by the following initial conditions
\be
a|_{t=0}=0
\,,\qquad
\dot{a}|_{t=0}
=\lambda\om
=
4|\alpha|B^{2}\,\f{\sqrt{k}}{1+k}
\,,
\ee
so that the mode $k$ is entirely determined by the initial speed (or, more precisely, the speed at vanishing area difference $a=0$, i.e.~$X_{1}=X_{2}$).
Let us recall the geometrical expression for the area difference velocity:
\be
\dot{a}=i(\balpha\bE-\alpha E)
\,,\quad
E=\la z_{1}|z_{2}\ra\la w_{1}|w_{2}\ra
\,,\quad
|E|^{2}=(X^{2}-a^{2})(Y^{2}-a^{2})
\,,\quad
|E|\Big{|}_{a=0}=XY
\,.
\ee
Since this implies that $|\dot{a}|\le 2|\alpha E|$, to be enforced at the initial time for which $a=0$, we obtain
\be
\f{2\sqrt{k}}{1+k}\le \f{2XY}{X^{2}+Y^{2}}
\,.
\ee
As illustrated on fig.\ref{plot:kfunction}, the left hand is a monotonically-increasing function of $k$, thus imposing a bound on the mode $k$, with $k_{max}\equiv \textrm{min}(X,Y)/\textrm{max}(X,Y)$.
In particular, the upper limit $k=1$, for which we get the solitonic solution $\tanh$ of the non-linear Schr\"odinger equation, is only allowed if $X=Y$ on the external legs of the candy graph.

\begin{figure}[!h]
    \centering
    \begin{subfigure}{0.32\textwidth}
        \centering
        \includegraphics[width=0.9\linewidth]{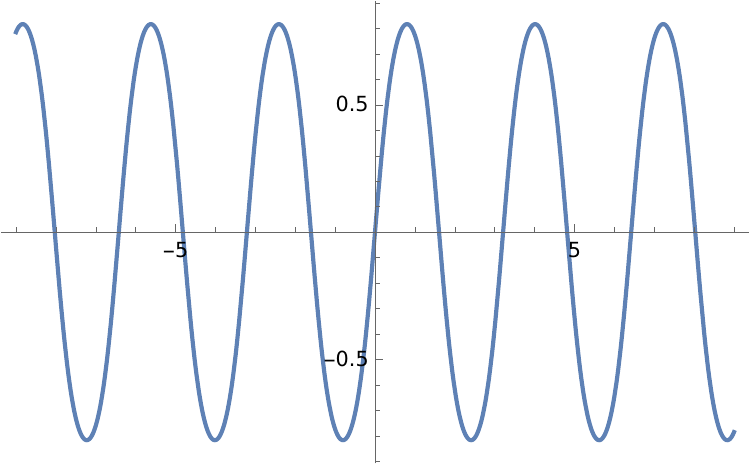}
        \caption{\small $B=1$, $k=0.5$}
    \end{subfigure}
    \begin{subfigure}{0.32\textwidth}
        \centering
        \includegraphics[width=0.9\linewidth]{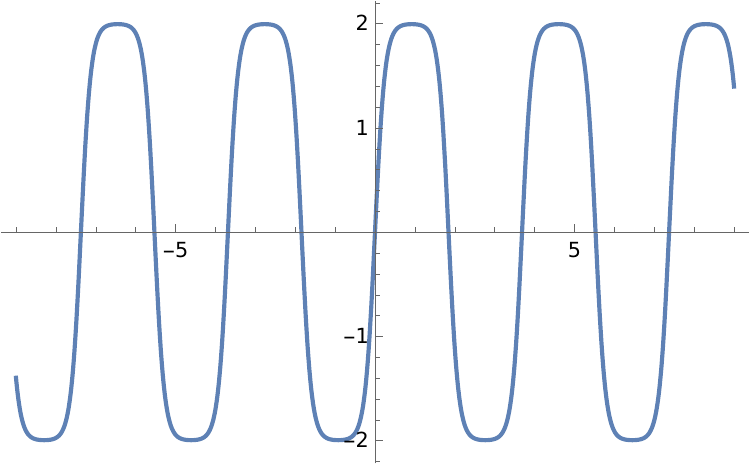}
        \caption{\small $B=2$, $k=0.99$}
    \end{subfigure}
        \begin{subfigure}{0.32\textwidth}
        \centering
        \includegraphics[width=0.9\linewidth]{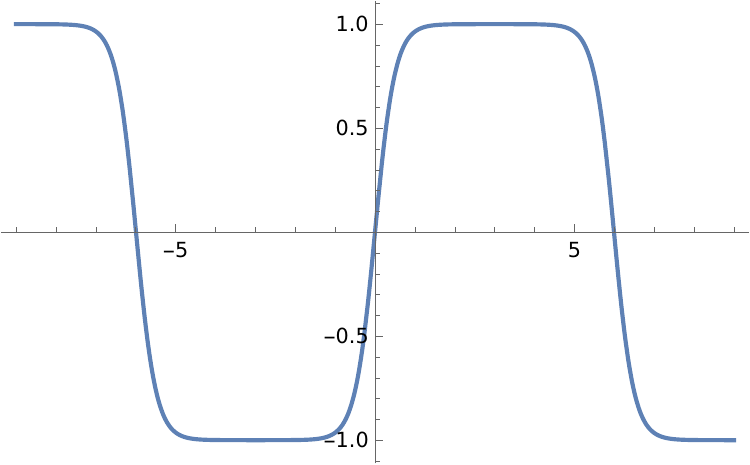}
        \caption{\small $B=1$, $k=0.9999$}
    \end{subfigure}
    \begin{subfigure}{0.32\textwidth}
        \centering
        \includegraphics[width=0.9\linewidth]{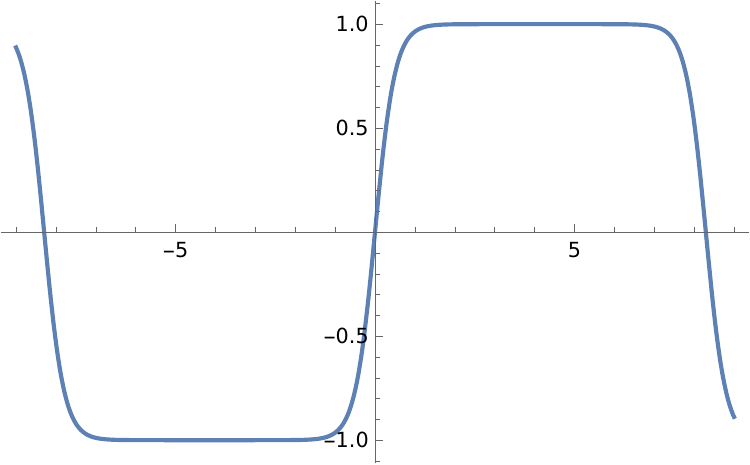}
        \caption{\small $B=1$, $k=0.999999$}
    \end{subfigure}
    \begin{subfigure}{0.32\textwidth}
        \centering
        \includegraphics[width=0.9\linewidth]{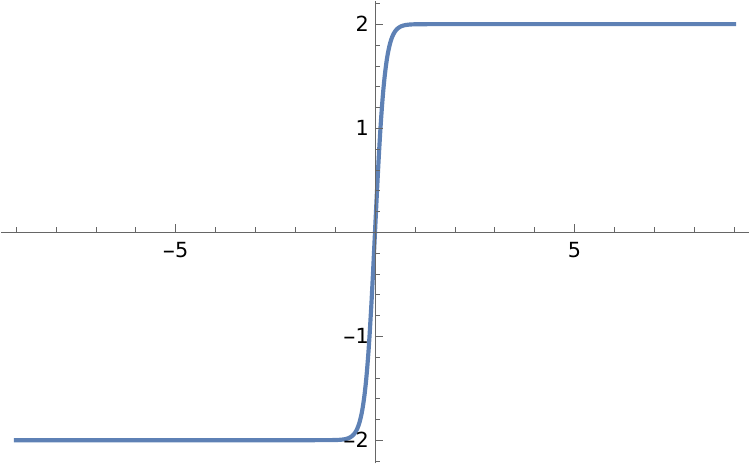}
        \caption{\small $B=2$, $k=1$}
    \end{subfigure}
    \begin{subfigure}{0.32\textwidth}
    	\centering
    	\includegraphics[width=0.9\linewidth]{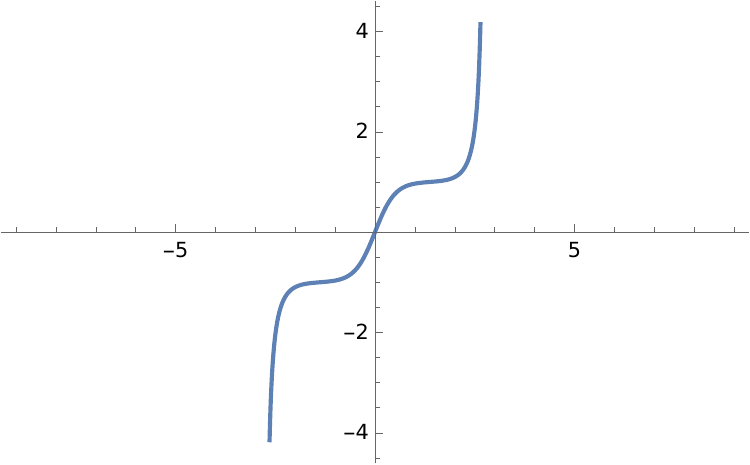}
    	\caption{\small $B=1$, $\dot{a}=2.001$}
    \end{subfigure}
    \caption{\small The area difference $a$ as a function of $t$ (on the horizontal axis) with various initial conditions and $\abs{\alpha}=1$, showing allowed trajectories with $\dot{a}<2B^2$ as in the plots (a) to (d), the critical trajectory with $k=1$ and $\dot{a}=2B^2$ as in the plot (e), and forbidden trajectories $\dot a > 2B^2$ as in the plot (f).}
    \label{plot:snsolutions}
\end{figure}

The period of the classical trajectory $a(t)$ is
\be
\cT
=
\f{4\tau_k}{\om}
=
\cI_{k}\left(\f\pi2\right)\,\f{\sqrt{2(1+k)}}{|\alpha|B}
\,,\qquad
\cT\underset{k\rightarrow0}\rightarrow
\f{2\pi}{2\sqrt{2}\,|\alpha|B}
\,,\quad
\cT\underset{k\rightarrow1}\rightarrow
+\infty
\,,
\ee
with a minimal period when the mode number $k$ vanishes and a period which stretches to infinity as $k$ goes towards its upper limit $k\rightarrow 1$.
We illustrate these oscillatory trajectories induced by the dynamics on the candy graph in fig.\ref{plot:snsolutions}.

Note that when the Hamiltonian coupling $\beta$ vanishes, then we have a purely oscillatory regime, with non-linear oscillations of the area difference $a$ while the total area $A$ remains fixed.

%%%%%%%%
\subsection{Hyperbolic Trajectories for the total area}
%%%%%%%%

We now turn to the dynamics of the total area $A$ driven by the same differential equation:
\be
\ddot{A}
=
8 |\beta|^{2}A (A^{2}-B^{2})
\,.
\ee
However, while the area difference $a$ was bounded in absolute value, the total area is only bounded from below by $\textrm{max}(X,Y)$, it is strictly positive and can grow arbitrarily large. Thus, it cannot be described by the $\sn$-solution, but rather by the other solution given by $1/\sn$  as mentioned earlier.
To arrive at this solution, we can mimic the previous construction, by introducing a modified Jacobi amplitude using a hyperbolic cosh-function instead of a trigonometric sine-function,
\be
\tcI_{\ell}(\xi)=\int_{0}^{\xi}\f{\rd \zeta}{\sqrt{\ell \cosh^{2}\zeta-1}}
\,,\qquad\textrm{with}\quad
\ell\ge 1
\,,
\ee
then defining a non-linearly deformed cosh-function:
\be
\chn_{\ell} \left(\tcI_{\ell}(\xi)\right) =\cosh \xi
\,,\qquad
\chn_{\ell} \Xi= \cosh \left(\tcI_{\ell}^{-1}(\Xi)\right)
\,.
\ee
This function automatically has the expected range of values and satisfies the desired differential equation:
\be\label{diff.eq.chn}
\pp_{\Xi}^{2}\chn_{\ell} \Xi
=
-(1+\ell)\chn_{\ell} \Xi +2\ell \chn_{\ell} ^{3}\Xi
\,,\qquad
(\pp_{\Xi}\chn_{\ell} \Xi)^2
=
(\chn_{\ell}  \Xi-1)
\,(\ell \chn_{\ell}  \Xi-1) 
\,.
\ee
This modified Jacobi amplitude is in fact related to the standard Jacobi amplitude, as is easily shown using a mere change of variable $u=1/\cosh \zeta$:
\be
\tcI_{\ell}(\xi)
=
\int_{\f1{\cosh\xi}}^{1}
\f{\rd u}{\sqrt{\ell}\,\sqrt{1-u^{2}}\sqrt{1-\f{u^{2}}{\ell}}}
\,,\quad\textrm{thus}\quad
\cI_{k}(\theta)+\sqrt{\ell}\ \tcI_{\ell}(\xi)=\tau_k
\,,\,\,\textrm{if}\,\,
\ell=\f1k
\,\,\textrm{and}\,\,
\sin\theta=\f1{\cosh \xi}
\,,
\ee
which leads to
\be
\chn_{\ell}\Xi=\f1{\sn_{k}\Theta}
\,,\qquad\textrm{for}\quad
\ell=\f1k
\,\,\textrm{and}\,\,
\Theta+\sqrt{\ell}\ \Xi =\tau_k
\,,
\ee
where we should remember that $\tau_k=\cI_{k}(\f\pi2)$ depends on the mode number $k$. So, as announced, this non-linearly modified cosh-function is simply the function $1/\sn$ (usually denoted as $\textrm{ns}$).

\begin{figure}[!h]
	\centering
	\begin{subfigure}{0.3\textwidth}
		\centering
		\includegraphics[width=0.9\linewidth]{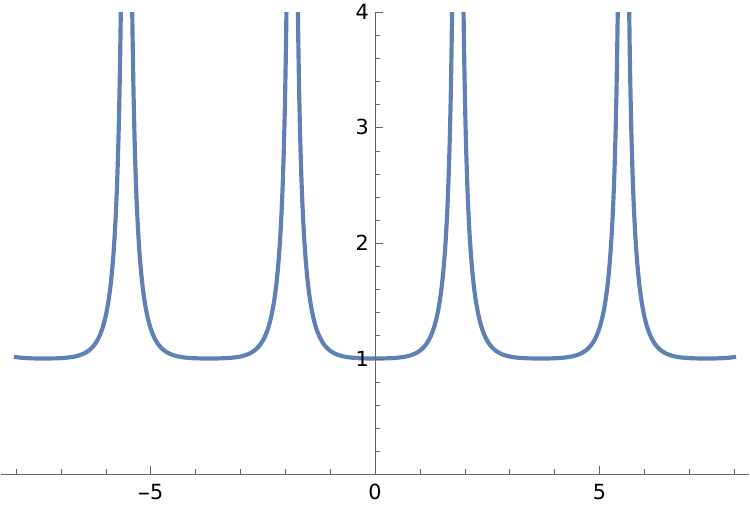}
		\caption{\small $B=1$, $\ell=1.01$}
	\end{subfigure}
	\begin{subfigure}{0.3\textwidth}
		\centering
		\includegraphics[width=0.9\linewidth]{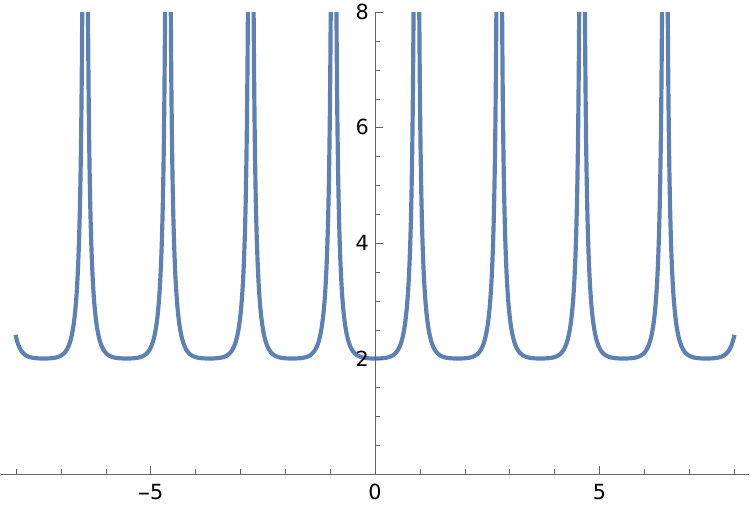}
		\caption{\small $B=2$, $\ell=1.01$}
	\end{subfigure}
	\begin{subfigure}{0.3\textwidth}
		\centering
		\includegraphics[width=0.9\linewidth]{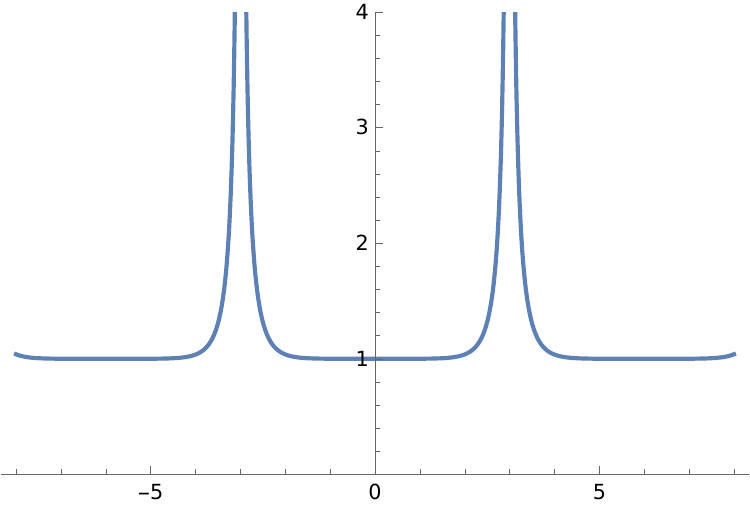}
		\caption{\small $B=1$, $\ell=1.0001$}
	\end{subfigure}
	\begin{subfigure}{0.3\textwidth}
		\centering
		\includegraphics[width=0.9\linewidth]{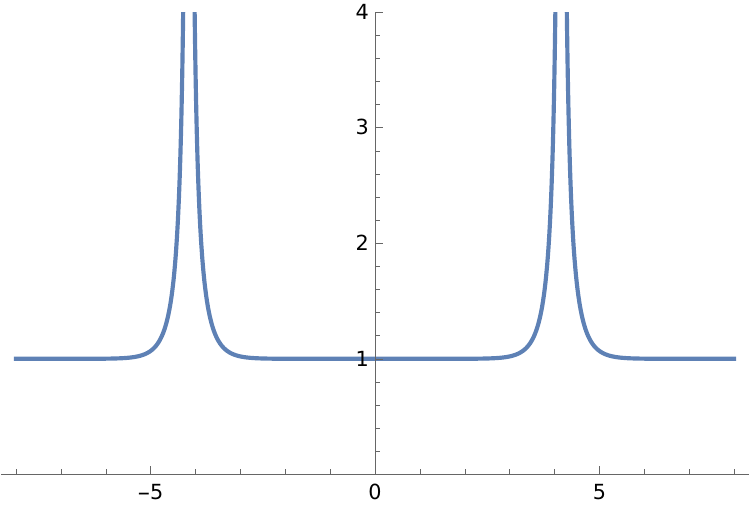}
		\caption{\small $B=1$, $\ell=1.000001$}
	\end{subfigure}
	\begin{subfigure}{0.3\textwidth}
		\centering
		\includegraphics[width=0.9\linewidth]{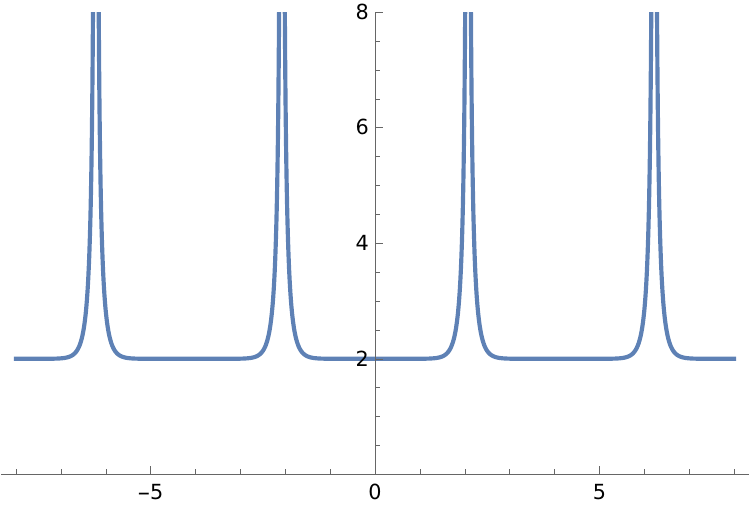}
		\caption{\small $B=2$, $\ell=1.000001$}
	\end{subfigure}
	\begin{subfigure}{0.3\textwidth}
		\centering
		\includegraphics[width=0.9\linewidth]{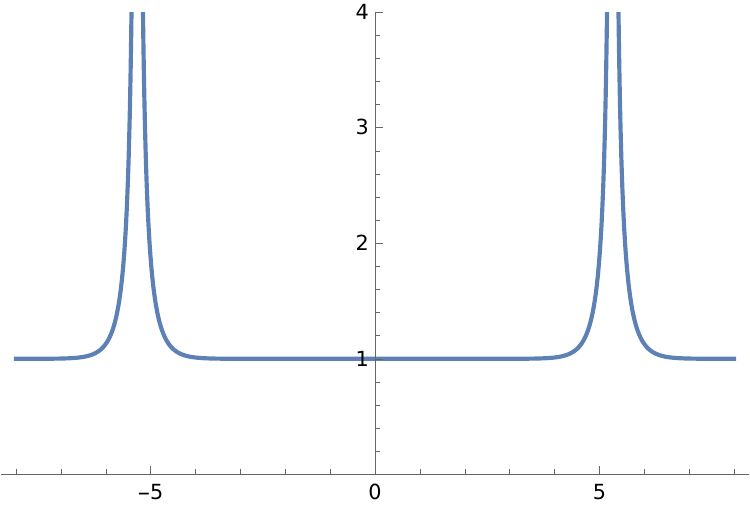}
		\caption{\small $B=1$, $\ell=1.00000001$}
	\end{subfigure}
    \caption{\small The total area $A$ as a function of $t$ (on the horizontal axis) with various initial conditions and $\abs{\beta}=1$.}
    \label{plot:chnsolutions}
\end{figure}

This result allows to integrate exactly the motion of the total area, namely
\be
A(t)=\pm \Lambda \,\chn_{\ell}(\Omega t)
\,,\qquad\,\textrm{with}\quad
\Omega
=
2|\beta|B\sqrt{\f2{1+\ell}}
\quad\textrm{and}\quad
\Lambda
=
\f{\Omega \ell}{2|\beta|}
=
B\sqrt{\f{2\ell}{1+\ell}} \geq 0
\,.
\ee
The sign $\pm$ is present in the expression of $A$ to guarantee its positivity: both functions $\chn_\ell$ and $-\chn_{\ell}$ are solutions of the second order differential equation in \eqref{diff.eq.chn}, however, the $\chn_\ell$ has a different (but constant) sign in different bounded domains of $t$; therefore, the solution to the evolution equation of $A$ depends on the chosen domain of $t$, and it is given by $\Lambda \,\chn_{\ell}(\Omega t)$ when $\chn_{\ell}(\Omega t) \geq 0$ and $- \Lambda \,\chn_{\ell}(\Omega t)$ when $\chn_{\ell}(\Omega t) \leq 0$.
We depict such expanding trajectories on fig.\ref{plot:chnsolutions}.
These solutions illustrate the hyperbolic behavior of the dynamics on the candy graph, where the total area diverges at the points $t = t_n^* = (2n+1)\tau_k/ (\sqrt{\ell} \Omega)$, $n\in \mathbbm{Z}$. This means that the total area always diverges in finite time, apart from the constant solution $A=B$, which is unstable. The different bounded domains of evolution are dynamically distinct, characterized by the point where the initial conditions are set, and they are separated by divergences at the points $t_n^* $. These divergences in the bulk area while the boundary area remains fixed at a finite value clearly signals a singularity. Such singularities are, in fact, the fluctuations that one wishes to renormalize away from the theory.

It follows that when the Hamiltonian coupling $\alpha$ vanishes, we obtain an accelerated regime for the total area $A$, similar to a bouncing cosmology transitioning from contraction to expansion, with fixed area difference $a$.
When we turn on both $\alpha$ and $\beta$, the dynamics of $a$ and $A$ remain decoupled and we get oscillations of the area difference on top of the hyperbolic evolution of the total area.

Let us underline that the divergence in the total area $A$ (in a region of fixed finite boundary area) is a priori different from the usual setting in loop quantum gravity, where a classical area going to 0 leads to an infinite curvature. Nevertheless, it is clear that, here too, our divergent area corresponds to a very large curvature regime. Indeed, as $A$ grows to $+\infty$, the holonomy around the loop $\tr\, g_2^{-1} g_1$ goes to 0 (instead of 2 for a trivial holonomy), corresponding to a maximal curvature for the Ashtekar-Barbero connection.

We see different (non mutually exclusive) possibilities to deal with the divergences. For instance, the coupling $\beta$ might have to be suppressed, so as to avoid those finite-time divergences. Also, introducing higher order terms in the Hamiltonian might tame the divergences or push them to infinite time. Or perhaps working with appropriate boundary conditions, such as a suitable time-dependent function $B(t)$, might lead to keeping the divergences under control, as we discuss in the next section.
Another possibility is that it is simply a question of choosing the correct notion of time.
Indeed, our choice of the time coordinate $t$ is simply a choice of lapse function. In the absence of a natural choice of observer or proper time, we could easily change our gauge choice for the lapse and consider another time coordinate $\ntau$ differing from $t$ by a field-dependent factor, for instance an arbitrary power of the volume density. In our simplified framework, this would amount to introducing factors of the bulk area $A$ in the time coordinate, namely
\be\label{rescaled_time}
\rd \ntau= A^{s} \rd t
\,,
\ee
for some exponent $s>0$.
Then the question is whether such a choice of new time coordinate might push, or not, the observed divergences to infinity.

\begin{figure}[!ht]
	\centering
	\includegraphics[height=40mm]{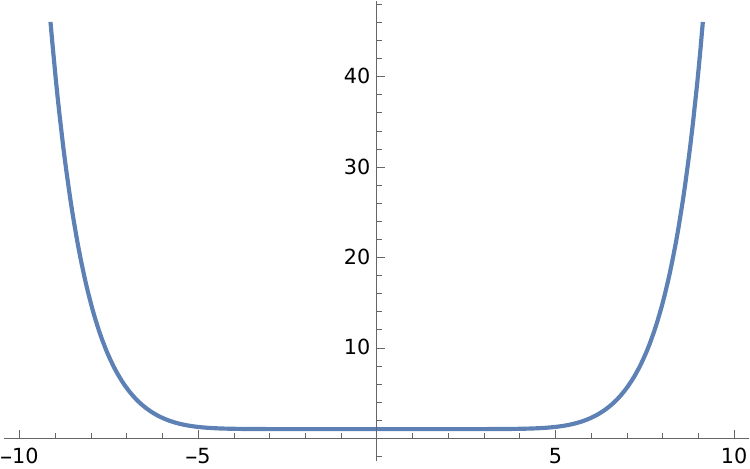}
	\caption{\small Plot of the total area $A$ as a function of the rescaled time $\rho$ (on the horizontal axis) for the values $(B, \ell) = (1, 1.0001)$ and $\abs{\beta}=1$.}
	\label{plot.A(tau)}
\end{figure}

Let us demonstrate how this idea works in the $s=1$ case, that is the simplest choice $\rd \ntau= A\rd t$.
The integration of this equation yields
\begin{equation}
\ntau(t)  = \frac{1}{2\abs{\beta}} \arctanh\left[ \sn_{\ell^{-1}} (\sqrt{\ell} \Omega t) \right]
\end{equation}
As $t$ approaches a point $t^*$ where $A$ diverges, the function $\sn$ goes to $\pm 1$, leading to an asymptotic behavior for $\ntau$:
\begin{equation}
\ntau(t) 
\underset{t\rightarrow t^*}\approx
-\frac{\sgn(t^{*}-t)}{2\abs{\beta}} \,\ln |t^* - t|\,
\end{equation}
This controlled divergence of the rescaled time parameter $\ntau$ pushes the divergence of the total area $A$ to infinite time $\ntau$ as illustrated in fig.\ref{plot.A(tau)}.
Indeed, as $t$ approaches $t^*$ from below (above) and $|t^* - t|<1$, $\ntau$ approaches $+ \infty$ ($- \infty$ respectively) and $\sgn(u) = \sgn(t^{*}-t)$, we then obtain
\begin{equation}
	A(t) = \frac{d\ntau}{dt} \underset{t\rightarrow t^*}\approx \frac{1}{2\abs{\beta}\abs{t^* -  t}}\quad , \quad 
	\abs{t^* -  t} \underset{t\rightarrow t^*}\approx e^{-2\abs{\beta}\abs{\ntau}} \qquad 
	\Rightarrow \qquad A(\ntau) \underset{t\rightarrow t^*}\approx \frac{1}{2\abs{\beta}} e^{2\abs{\beta}\abs{\ntau}}.
\end{equation}
In fact, we can proceed with a more systematic analysis of the asymptotic regime of $A$ as a function of the time $\ntau$ defined via \eqref{rescaled_time}. Indeed, if we assume that the total area $A$ is very large, i.e.~$A\rightarrow+\infty$, then the differential equation driving its evolution simplifies in this asymptotic regime to approximatively
\be
\ddot{A}\approx 8|\beta|^{2}A^{3}
\,.
\ee
This implies a power law asymptotic behavior :
\be
A\underset{A\rightarrow+\infty}\approx\f{1}{2|\beta|\,|t^{*}-t|}
\,.
\ee
This $1/t$-divergent behavior allows for a simple integration for the new time coordinate. We can indeed compute, say for $t$ approaching $t^{*}$ from below, for $s>1$,
\begin{align}
&\textrm{for }s\ne1,\,\,
\ntau=-\int^{t^{*}}_{t}A^{s}\approx
\f1{(s-1)(2|\beta|)^{s}\,(t^{*}-t)^{s-1}}
\propto
A^{s-1}
\quad\Rightarrow\quad
A\propto \ntau^{\f1{s-1}}
\,,\\[1em]
&\textrm{for }s=1,\,\,
\ntau=-\int^{t^{*}}_{t}A\approx
-\f1{2|\beta|}\ln(t^{*}-t)
\quad\Rightarrow\quad
A\propto e^{{2|\beta|}\ntau}
\,.
\end{align}
This simple analysis shows that, as soon as $s\ge 1$, the divergence of the total area $A$ in the new time $u$ is pushed from finite time to infinite time.

Overall, all the above mentioned approaches to handle the dynamical divergences in the bulk observables deserve further investigation, which we leave for future works.
Let us nevertheless point out that this type of bulk divergence is different from the typical singularity setting where one consider tiny volumes or areas. Here it is a divergent bulk area at fixed boundary area. Hence it is unlikely to be resolved at the quantum level by the discrete spectra of areas and volumes, and loop quantum gravity will have to address this issue, either by taming or renormalizing this divergent behavior of curvature excitations.

%%%%%%%%
\section{Dynamics Beyond the 2-legged Loop}\label{Beyond2leg}
%%%%%%%%

Up till now, our analysis assumed fixed boundary conditions, that is fixed areas $X$ and $Y$, and thus constant $B$.
In order to connect our 2-leg candy graph to an exterior, or in other words, extend the model to study the effect of evolving boundary conditions on the dynamics of the curvature excitation around the bulk loop, we would like to relax the condition of fixed $X$ and $Y$.

It turns out that it is not straightforward to have evolving $X$ and $Y$ area functions in a natural controlled way.
Indeed, a first idea would be to simply consider an arbitrary function $B(t)$ and plug it directly into our equations of motion for the total area $A$ and area difference $a$. However, one would like to have a Hamiltonian generating such modified equations of motion. This task faces a very obvious obstacle: our present Hamiltonian in \eqref{Hamiltonian.Ansatz} commutes with $\vX_1+\vX_2$, and thus irremediably leaves the area $X=|\vX_{1}+\vX_{2}|$ constant in time. To overcome this issue, we would need to incorporate $X$ in our phase space and identify its conjugate variable (or at least an appropriate variable not commuting with it). This would then allow us to look for suitable potential terms compatible with an evolving area $X$, which hopefully does not drastically modify the equations of motion studied previously.

Note that this method of enhancing the bulk phase space with a suitable boundary phase space to enable flexible and dynamical boundary conditions is consistent with the logic of gravitational edge modes \cite{Freidel:2019ees,Freidel:2020xyx,Freidel:2020svx,Freidel:2020ayo}, already understood as playing a key role in the gravitational memory effects, e.g.\cite{Strominger:2014pwa,Pasterski:2015tva,Strominger:2017zoo,Donnay:2018ckb,Compere:2018aar,Khera:2022mfe}, and meant to play a crucial role in quantum gravity, especially reformulated in terms of quantum reference frames \cite{Donnelly:2016auv,Carrozza:2022xut,Goeller:2022rsx,DeVuyst:2024khu}. We explore some ideas of including dynamical boundary degrees of freedom in the model in the following.

%%%%%%%%
\subsection{Dynamical boundaries}
%%%%%%%%

One can upgrade the bulk phase space and enhance it with a boundary phase space by including new boundary spinors $|z\ra$ and $|w\ra$ corresponding to the boundary fluxes $\vX$ and $\vY$, on top of the bulk spinors $|z_{i}\ra$ and $|w_{i}\ra$ used up to now. It turns out that the phase of the spinor $|z\ra$, respectively $|w\ra$, is now the canonical momentum for the area $X$, respectively $Y$.
It is instructive to see how this works. Let us focus on the spinor $z\in\C^{2}$ and its flux $\vX=\f12\la z|\vsigma| z\ra\in\R^{3}$. As mentioned earlier, the spinor contains one parameter on top of the flux, and it is simply its phase as stressed in \cite{Freidel:2009ck,Freidel:2010tt}. In fact, one can invert the mapping, following e.g. \cite{Livine:2013tsa}, as
\be
|z\ra
=
\mat{c}{z^{0} \\ z^{1}}
=
e^{i\vphi}\,\mat{c}{e^{-i\f\psi2}\sqrt{X+X^{z}}\\e^{i\f\psi2}\sqrt{X-X^{z}}}
\,,\quad\textrm{with}\quad
\tan \psi=\f{X^{y}}{X^{x}}
\,,\quad
e^{i\psi}
=\f{X^{x}+iX^{y}}{\sqrt{(X^{x})^{2}+(X^{y})^{2}}}
\,,
\ee
where both phases live in the same interval, $\vphi,\psi\in[-\pi,+\pi]$.
To explicitly compute the Poisson brackets, it is convenient to express all the quantities in terms of the spinor components,
\be
X^{x}+iX^{y}=\bz^{0}z^{1}
\,,\,\,
X^{z}=\f12(\bz^{0}z^{0}-\bz^{1}z^{1})
\,,\,\,
X=\f12(\bz^{0}z^{0}+\bz^{1}z^{1})
\ee
\be
e^{4i\vphi}=\f{z^{0}z^{1}}{\bz^{0}\bz^{1}}
\,,\,\,
e^{2i\psi}=\f{\bz^{0}z^{1}}{z^{0}\bz^{1}}
\,. \label{spinor_phase}
\ee
The canonical bracket $\{z^{A},\bz^{B}\}=-i\delta^{AB}$ then implies
\be
\{X,X^{z}\}= 0 =\{e^{4i\vphi},e^{2i\psi}\}
\,,\quad
\{X,e^{2i\psi}\}= 0 =\{X^{z},e^{4i\vphi}\}
\,,
\ee
\be
\{X,e^{4i\vphi}\}=2ie^{4i\vphi}
\,,\quad
\{X^{z},e^{2i\psi}\}=-2ie^{2i\psi}
\,,
\ee
from which we extract the canonical brackets:
\be
\{X,\vphi\}=\f12
\,,\quad
\{X^{z},\psi\}=-1
\,,\quad
\{X^{z},\vphi\}=0=\{X,\psi\}
\,.
\ee
We thus have two decoupled canonical flux-angle pairs: on the one hand, the $2\{X,\vphi\}=1$ pair with the norm $X$ and the global phase $\vphi$, and on the other hand, the $\{\psi,X^{z}\}=1$ with a single flux component $X^{z}$ and the relative phase $\psi$. 
This can be checked directly working the kinetic term encoding the symplectic structure:
\beq
-2X\pp_{t}\vphi+X^{z}\pp_{t}\psi
&=&
\f i4(\bz^{0}z^{0}+\bz^{1}z^{1})\pp_{t}\ln\f{z^{0}z^{1}}{\bz^{0}\bz^{1}}
-
\f i4(\bz^{0}z^{0}-\bz^{1}z^{1})\pp_{t}\ln\f{\bz^{0}z^{1}}{z^{0}\bz^{1}}
\\
&=&
\f i2(\bz^{0}\pp_{t}z_{0}-z^{0}\pp_{t}\bz_{0}+\bz^{1}\pp_{t}z_{1}-z^{1}\pp_{t}\bz_{1})
=
\f i2\big{(}
\la z|\pp_{t}z\ra-\la \pp_{t} z|z\ra
\big{)} 
\,,\nn
\eeq
where we recognize the first term $X\pp_{t}\vphi$ as the norm-phase conjugate pair, and the second term $X^{z}\pp_{t}\psi$ which defines the symplectic structure on the 2-sphere at fixed norm. The latter indeed encodes the $\su(2)$-brackets for the flux, $\{X^{a},X^{b}\}=\eps^{abc}X^{c}$, namely
\be
\left|
\begin{array}{c}
\{X^{z},X^{x}\}=+X^{y}
\,,\vspace*{2mm}\\
\{X^{z},X^{y}\}=-X^{x}\,,
\end{array}
\right.
\quad\Longleftrightarrow\quad
\left|
\begin{array}{l}
\Big{\{}X^{z},(X^{x})^{2}+(X^{y})^{2}\Big{\}}=0
\,,\vspace*{2mm}\\
\displaystyle
\{\psi,X^{z}\}=\f{1}{1+\tan^{2}\psi}\left\{\f{X^{y}}{X^{x}}\,,\,X^{z}\right\}=1\,.
\end{array}
\right.
\ee

\medskip

Now, having incorporated the boundary fluxes $X$ and $Y$ in the phase space through the spinors $z$ and $w$, we can introduce bulk-boundary matching constraints explicitly enforcing the Gauss laws $\vX=\vX_{1}+\vX_{2}$ and $\vY=\vY_{1}+\vY_{2}$, then we can exactly define a time-dependent Hamiltonian driving any desired evolution for $X(t)$, $Y(t)$ and thus $B(t)$. This is not completely obvious, but nevertheless possible. It is in fact very similar to driving the evolution of the center of mass of a two-particle system without affecting their interaction\footnote{
Starting with a system of two interacting particles, say in one dimension, defined by the action principle in Hamiltonian form with a translation-invariant potential $V$,
\be
S_\text{free}=\int \rd t\, \big{[}
p_{1}\pp_{t}x_{1} +p_{2}\pp_{t}x_{2}
- H
\big{]}
\quad\textrm{with}\quad
H=
\f{p_{1}^{2}}{2}+\f{p_{2}^{2}}{2}+V(x_{1}-x_{2})
\,,
\nn
\ee
the total momentum $p_{1}+p_{2}$ is obviously a constant of motion. Thus one cannot simply impose that $p_{1}+p_{2}=P(t)$, where $P(t)$ is an arbitrary function of our choosing. To effectively drive the center of mass in a consistent way, one needs to not only introduce the constraint with a Lagrange multiplier, but also introduce another Lagrange multiplier forcing the primary Lagrange multiplier to vanish and a new force term coupled to the center of mass position $\f12(x_{1}+x_{2})$, as follows:
\be
S_\text{forced}
=
\int \rd t\, \bigg{[}
p_{1}\dot{x}_{1} +p_{2}\dot{x}_{2}
-\left(
\f{p_{1}^{2}}{2}+\f{p_{2}^{2}}{2}+V(x_{1}-x_{2})
\right)
+\nu\mu
-P(t)\f12(\dot{x}_{1}+\dot{x}_{2})
+\mu(p_{1}+p_{2}-P(t))
\bigg{]}\,.
\nn
\ee
The equations of motion then read, with $x\equiv x_{1}-x_{2}$:
\be
\left|
\begin{array}{l}
\dot{x}_{1}+\mu=p_{1}\,,
\\
\dot{x}_{2}+\mu=p_{2}\,,
\end{array}
\right.
\quad
\left|
\begin{array}{l}
\dot{p}_{1}=V'(x)+\f12\dot{P}
\,,\\
\dot{p}_{2}=-V'(x)+\f12\dot{P}
\,,
\end{array}
\right.
\quad
\left|
\begin{array}{l}
{\mu}=0
\,,\\
\nu+p_{1}+p_{2}=P(t)
\,,
\end{array}
\right.
\ee
implying that $\nu$ is a constant, $\nu=\nu_{0}$, and thus $p_{1}+p_{2}=P(t)-\nu_{0}$ for the center of mass and $\ddot{x}=2V'(x)$ for the relative motion, as desired.}.

Instead of pushing the possibility of putting by hand an arbitrary evolution for $B(t)$, we prefer to focus here on investigating the possibility of endowing the boundary degrees of freedom with natural Hamiltonian dynamics.
This is relevant for two reasons:
first, in order to create a template for the dynamics of curvature fluctuations in loop quantum gravity, we should start by analyzing the simplest Hamiltonian for the boundary data and see how it affects the bulk dynamics of the curvature around the loop;
second, the evolution of the geometry of the boundary data should, at the end of the day, reflect the evolution of the exterior geometry, i.e.~the dynamics of the rest of the spin network. Therefore, we expect the boundary Hamiltonian to be similar in construction to the bulk Hamiltonian, and at least at leading order, the exterior and bulk evolve in a similar fashion.

Thus, we can consider a total Hamiltonian of the form
\be
\bfH=H+H_{X}+H_{Y}+\vmu\cdot(\vX_{1}+\vX_{2}-\vX)+\vnu\cdot(\vY_{1}+\vY_{2}-\vY)
\,,
\ee
with the bulk Hamiltonian $H$ given by \eqref{Hamiltonian.Ansatz} and studied in the previous sections, new boundary Hamiltonians inducing potentially non-trivial dynamics for the boundary fluxes, and the two Gauss law constraints $\vX_{1}+\vX_{2}=\vX$ and $\vY_{1}+\vY_{2}=\vY$, enforced by the Lagrange multipliers $\vmu$ and $\vnu$.
Therefore, we assume that the bulk is only coupled to the boundary through the Gauss laws and that the boundary Hamiltonians $H_{X}+H_{Y}$ drive the bulk dynamics without back-reaction.

Let us then discuss the dynamics of the flux $\vX$ induced by $H_{X}$, the same discussion applies to the flux $\vY$.
Having the canonical pair $(X,\vphi)$ at hand, one can manufacture any Hamiltonian for the area $X$.
We should nevertheless keep in mind that $\vphi$ is an angle, defined as a phase from the spinor components as in \eqref{spinor_phase}.
This suggests two simple ansatz for the boundary Hamiltonian:
\begin{itemize}
\item $H_{X}=X^{2}+r\cos4\vphi$, with constant parameter $r\in\R$:
this is the equivalent of an anharmonic oscillator. It is actually the Hamiltonian of a Josephson junction in spintronics and quantum circuits.
The equations of motion are:
\be
\dot{X} = \{X,H_{X}\} = -2r\sin4\vphi
\,,\quad
\dot{\vphi}=\{\vphi,H_{X}\}=X
\quad\Rightarrow\quad
\ddot{X}=8X(X^{2}-H_{X})\,,
\ee
where $H_{X}$ is now obviously a constant of motion. This is exactly the same equation as the bulk areas and supports the naturalness of those non-linear differential equations. Nevertheless, if we forbid oscillating solutions which would lead to negative values of the boundary area $X$, we are left with  trajectories diverging in finite time, which we would probably prefer to avoid.

\item $H_{X}=r(z^{0}z^{1}+\bz^{0}\bz^{1})$ with constant parameter $r>0$:
this is probably the most natural boundary Hamiltonian, since it is polynomial in the spinor components, where we recognize a boost generator acting in the spinor phase space \cite{Dupuis:2011wy}. The equations of motion for $X$ are then
\be
\dot{X}=\{X,H_{X}\}=ir(z^{0}z^{1}-\bz^{0}\bz^{1})
\,,\quad
\ddot{X}= 4r^{2}X
\,,
\ee
which are simply integrated as exponentials $e^{\pm 2rt}$ (or hyperbolic functions in $\cosh(2rt)$ and $\sinh(2rt)$).
Endowing both $X$ and $Y$ with this dynamics leads to an evolving $B(t)^{2}=(X^{2}+Y^{2})/2$ function. Plugging this in the equation of motion for the total bulk area \eqref{A_eom} seems to lead to exponential solutions $A^{2}=B^{2}+constant$ if the evolution of $A$ is synchronized with the frequency $r$. This points to a possible resolution of the divergence in finite time driven by the boundary dynamics, and it should be further investigated in a systematic way.

\end{itemize}

%%%%%%%%
\subsection{Coupled tetrahedra}
%%%%%%%%

\begin{figure}[!h]
    \centering
    \includegraphics[width=120mm]{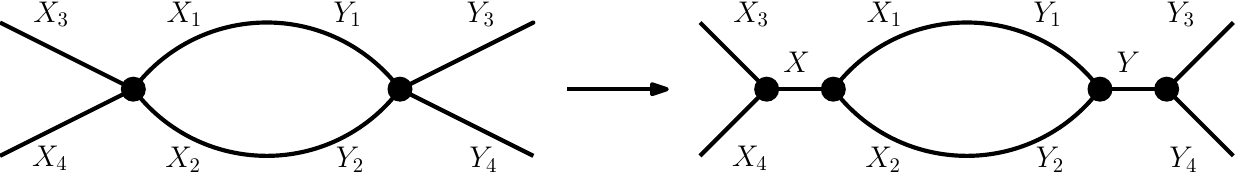}
    \caption{\small Unfolding the 4-valent nodes of the candy graph}
    \label{fig:unfoldedcandygraph}
\end{figure}

Another possible path to include dynamical boundary degrees of freedom is to go back to the original setting of 4-valent intertwtiners, i.e.~studying the candy graph made of a single loop connecting two 4-valent nodes, as drawn on fig.\ref{fig:unfoldedcandygraph}. Now the boundary fluxes will be $\vX_{3,4}$ and $\vY_{3,4}$, while $\vX$ and $\vY$ become bulk fluxes. They are actually internal to the two 4-valent nodes, when unfolding them into pairs of 3-valent nodes by writing $\vX=\vX_{1}+\vX_{2}=-(\vX_{3}+\vX_{4})$ and $\vY=\vY_{1}+\vY_{2}=-(\vY_{3}+\vY_{4})$.

\medskip

The phase space associated to a 4-valent node has been extensively studied in the loop quantum gravity literature, and it has been shown that a 4-valent node is dual to the geometry of a tetrahedron.
Considering the phase space $(\C^{2})^{\times 4}$ formed by the spinors $z_{1},..,z_{4}$, each spinor decomposes into a phase plus a 3d flux $\vX_{e}$ for $e\in\{1,2,3,4\}$. 
The four spinor phases can be identified with 2d orthonormal frames attached to each triangle \cite{Freidel:2009ck}, and they allow to define a phase space for the tetrahedron geometry in which the four triangle areas can vary \cite{Livine:2013tsa}.
The Gauss law constraints $\sum_{e}\vX_{e}=0$ generate the gauge invariance of the spinors under $\SU(2)$ transformation, i.e.~3d rotations of the flux vectors. The symplectic reduction by this set of constraints reduces the 16-dimensional phase space $(\C^{2})^{\times 4}$  to a $\SU(2)$-invariant phase space of dimension $16-2\times 3=10$. This reduced phase space can be parametrized by four spinor phases, which are canonically conjugate to the four flux norms $X_{e}$. 
This leaves us with 2 parameters, representing the true internal degrees of freedom of the node, which we can identify as the internal area $X=|\vX_{1}+\vX_{2}|=|\vX_{3}+\vX_{4}|$ and the internal dihedral angle $\vphi_X$ between the planes $(\vX_{1},\vX_{2})$ and $(\vX_{3},\vX_{4})$:
\be
\cos\vphi=\f{(\vX_{1}\w \vX_{2})\cdot (\vX_{3}\w \vX_{4})}{|\vX_{1}\w \vX_{2}|\,|\vX_{3}\w \vX_{4}|}
\,,\qquad
\{\vphi_X\,,\,X\}=1
\,,
\ee
as can be checked directly\footnotemark{}.
\footnotetext{
Taking into account that $|\vX_{1}+\vX_{2}|^{2}=X_{1}^{2}+X_{2}^{2}+2\vX_{1}\cdot\vX_{2}$ and $|\vX_{1}\w\vX_{2}|^{2}=X_{1}^{2}X_{2}^{2}-(\vX_{1}\cdot\vX_{2})^{2}$ commute with each other and with the norms $X_{1}$ and $X_{2}$, we can compute:
\beq
\{\vX_{1}\cdot\vX_{2}\,,\, (\vX_{1}\w \vX_{2})\cdot (\vX_{3}\w \vX_{4})\}
&=&
\vX_{1}\cdot(\vX_{2}\w(\vX_{2}\w(\vX_{3}\w\vX_{4})))-\vX_{2}\cdot(\vX_{1}\w(\vX_{1}\w(\vX_{3}\w\vX_{4})))
\nn\\
&=&
(\vX_{1}\w\vX_{2})\cdot((\vX_{1}+\vX_{2})\w(\vX_{3}\w\vX_{4}))
=
(\vX_{1}+\vX_{2})\cdot((\vX_{1}\w\vX_{2})\w(\vX_{3}\w\vX_{4}))
\nn\\
&=&
|\vX_{1}+\vX_{2}|\,\big{|}(\vX_{1}\w\vX_{2})\w(\vX_{3}\w\vX_{4})\big{|}
\,,\nn
\eeq
where we used the simple fact that $\vX_{1}+\vX_{2}=-(\vX_{3}+\vX_{4})$ is orthogonal to both $\vX_{1}\w\vX_{2}$ and $\vX_{3}\w\vX_{4}$. It is then straightforward to obtain the canonical Poisson bracket $\{\vphi\,,\,X\}=1$ from this.
}
The geometrical interpretation is obtained by reconstructing the dual tetrahedron, such that the four fluxes $\vX_{e}$ are identified as the (outward oriented) normal vectors to its four triangles. Up to 3d rotations, a tetrahedron is defined by 6 parameters, usually taken as its six edge lengths. Here, instead, we consider the area of its four triangles, identified as the flux norms $X_{e}$, plus the internal area $X=|\vX_{1}+\vX_{2}|$ and its conjugate internal dihedral angle $\vphi_X$ (see e.g. \cite{Bianchi:2011ub,Bianchi:2012wb,Hedeman:2014iki}).
The 2d phase space parametrized by $X$ and $\vphi$, also called the Millson-Kapovich phase space \cite{KapovichMillson}, is topologically a 2-sphere and it encodes, up to 3d rotations, the degrees of freedom of the tetrahedron geometry at fixed triangle areas.
To characterize the geometry of the tetrahedron, we could in principle work with more usual variables, such as the (oriented) squared volume $U\equiv \vX_{1}\cdot(\vX_{2}\w\vX_{3})$ and the quadrupole moment $T_{ab}\equiv \sum_{e}X_{e}^{a}X_{e}^{b}$ probing the shape of the tetrahedron \cite{Goeller:2018jaj}. However, using the variables $X$ and $\vphi$ which are canonical variables, allows to define and analyze the Hamiltonian dynamics in a simple manner.

Having the canonical pairs $(X,\vphi_X)$ and $(Y,\vphi_Y)$ as the internal degrees of freedom of the two 4-valent nodes, we can use them to write down new Hamiltonian terms which will make the areas $X$ and $Y$ evolve, hopefully without affecting too much the dynamics of the areas $X_{1,2}$ and $Y_{1,2}$ as well as $X_{3,4}$ and $Y_{3,4}$.
One would write an extended ansatz $\bfH=H+H_{X}+H_{Y}$, with our original Hamiltonian $H$ in \eqref{Hamiltonian.Ansatz} and new Hamiltonian terms for the two internal sectors $(X,\vphi_X)$ and $(Y,\vphi_Y)$.
For instance, one could imagine to endow the internal sectors with the dynamics of harmonic oscillators (with a $\vphi^{2}$ potential), or even more appropriately by the dynamics of Josephson junctions as in quantum circuits (with a $\cos\vphi$ potential).
One would hope to get decoupled oscillations of $X$ and $Y$ within each node, which would then drive the evolution of the total area and areas difference around the bulk loop according to our differential equations \eqref{A_eom} and \eqref{a_eom} with evolving ``boundary'' data $B(t)$.
However, while the internal norms $X$ and $Y$ obviously commute with the original bulk Hamiltonian $H$, the internal angles $\vphi$ and $\tvphi$ do not.
We then should derive and solve the new coupled dynamics, or see if it is possible to find a suitable redefinition of the internal angles which commutes with $H$. We postpone this more complex analysis to future investigations.

With the approach above, we now have a general template of a (classical) Hamiltonian driving the dynamics of both nodes and links (i.e.~for both intertwiners and spins) in the candy graph model.
Nevertheless, one final ingredient seems to be missing: spin exchange across nodes, for instance between the fluxes around the bulk loop and the boundary fluxes, such as between $\vX_{1,2}$ and $\vX_{3,4}$. This would complete our candy graph kit with all possible dynamical features, and make it fully ready to be connected to other spin network blocks and interact with them.
In the next section, we go over few ideas on how to possibly construct more complex and interacting structures using the candy graph as a basic block.

%%%%%%%%
\subsection{Spin network architecture and traveling waves}\label{sec:architecture}
%%%%%%%%

\begin{figure}[!h]
    \centering
        \includegraphics[width=130mm]{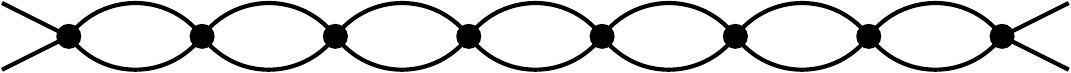}
    \caption{\small The coil chain or garland: to study wave propagation in LQG}
    \label{fig:coilchain}
\end{figure}

In order to move from the candy graph to more complex spin networks, we can consider gluing together candy graphs to form a larger spin network graph. Since the basic candy graph has four external legs, we could imagine crystal structures (like diamond lattices) or water-like structures, depending on how the candy graphs connect to each other, then study the possible dynamics and explore the corresponding phases. We would like to start with the simplest way of gluing candy graphs, namely along a one-dimensional coil chain or garland as in fig.\ref{fig:coilchain}. 
Such a simple architecture is enough to study the propagation of geometry fluctuations and the possible emergence of traveling waves, which could reproduce semi-classical gravitational waves or reveal new purely quantum modes in loop quantum gravity.

To define a Hamiltonian dynamics for this architecture, we can use what we explored for the candy graph model. Namely, we can unfold every node and stretch them to make the internal link apparent as we did for the 2-leg candy graph model. If one uses the Hamiltonian ansatz we studied in the previous section, while keeping the ``boundary'' areas fixed, one can see that all the loops remain decoupled. That is, we produce localized fluctuations and evolutions, but no traveling waves.

\begin{figure}[!h]
	\centering
	\includegraphics[width=170mm]{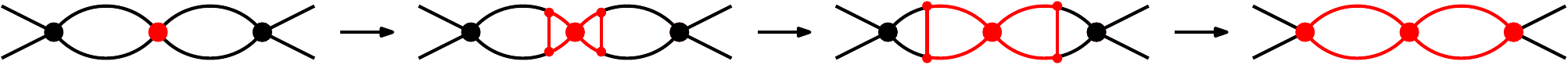}
	\caption{\small Depicting the generation of a figure-8 excitation via a traveling entangled spin excitation à la Thiemann-Varadarajan.}
	\label{fig:travellingexcitation}
\end{figure}

To go beyond this elementary template dynamics, we should allow the areas of the internal links to evolve and area excitations to travel from one loop to the next by crossing through nodes.
We envision two ways to realize this:
\begin{itemize}
\item we can couple the internal degrees of freedom of neighboring nodes, that is the $(X,\vphi_X)$ variables. A first step in this direction would be to include the discretized 3-curvature terms $H_L$ in \eqref{HL}, which we constructed in section \ref{I.C} and were obtained based on the regularized version of the 3-curvature (Ricci scalar) in LQG \cite{Alesci:2014aza}. Indeed, while the internal areas $X$ would commute with the term $H_L$ (since all areas commute with $H_L$), the angles $\vphi_X$ would not. Another possibility is to look at the volume-volume interaction, as proposed in \cite{Feller:2015yta}.
\begin{figure}[!ht]
	\centering
	\includegraphics[height=8mm]{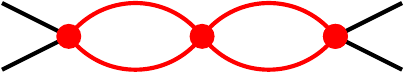}
	\caption{\small Hamiltonian interaction term with figure-8 excitations of the holonomy around a double loop.}
	\label{fig:8excitation}
\end{figure}

\item we could use holonomies going through nodes. Since the holonomies need to live on closed loop to ensure their gauge invariance, the simplest such loop is the figure-8 loop connecting three consecutive nodes, as shown on fig.\ref{fig:8excitation}.
This is obviously the correlated action of two neighboring elementary loops, which will become at the quantum level the creation/annihilation of entangled curvature excitations around those two loops.
In fact, one can view such a term as an effective interaction emerging from Thiemann-type Hamiltonians (see \cite{Thiemann:1996aw,Thiemann:1996av} for the action of Thiemann's Hamiltonian in loop quantum gravity), where one focuses on the creation of special transverse links at the nodes. Namely, one would create two such transverse links around the central node by the action of the Hamiltonian, have those transverse links travel to the extremities (which is technically realized via the action of a finite spatial diffeomorphism) of the figure-8, and finally have them re-absorbed by the nodes at the extremities, as drawn on fig.\ref{fig:travellingexcitation}.
This would naturally produce waves in the spirit of Thiemann-Varadarajan proposal \cite{Thiemann:2021hpa}.
\end{itemize}

\begin{figure}[!ht]
    \centering
        \includegraphics[width=150mm]{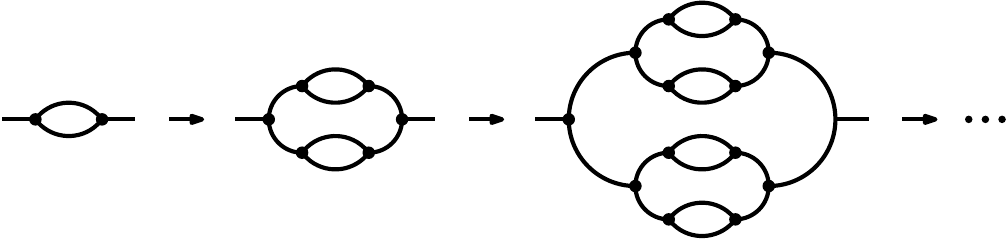}
    \caption{\small Melonic refinement: another setting to study renormalization of spin network edges accounting for local excitations of the curvature.}
    \label{fig:melonic}
\end{figure}

Finally, we would like to underline another interesting simple spin network architecture, beyond a 1d structure, which deserves to be analyzed in detail in future works. Namely, in the logic of renormalization in quantum field theory, building on the similarities\footnotemark{} between spin networks and Feynman diagrams, we can look at how excitations of curvature can develop on spin network links following the ideas of spin network coarse-graining of \cite{Charles:2016xwc,Livine:2019cvi}, in which one allows links to carry local physical degrees of freedom. 
This is done by inserting a loop along the link, i.e.~replacing the link with our 2-leg candy graph. Iterating this procedure yields a ``melonic'' refinement of the link, following the terminology developed in group field theory and tensor models (see e.g. \cite{Bonzom:2011zz,Lionni:2017yvi,Itoyama:2017emp,Bonzom:2019kxi}), as illustrated on fig.\ref{fig:melonic}. We can expect that studying the dynamics of successive such refinement would provide essential insights into the renormalization group flow of loop quantum gravity dynamics.

%\newpage
%%%%%%%%
\section{Outlook}
%%%%%%%%

In summary, in pursuit of a better understanding of the dynamics of the quantum geometry in Loop Quantum Gravity, we decided to focus on the simplest representative building block of spin network states: the {\it candy graph}, which consists of two nodes with links between them and an arbitrary number of open edges. Each node represent an elementary 3d volume. Curvature can develop along the loops formed by the links between the two nodes, while open edges define the boundary of the system. Gluing such candy graphs together allows to build scalable spin network architectures.

To capture the essential features of the dynamics on the candy graphs, we reduce the system to its simplest configuration: a single loop with two open links.  This is basically a Wilson loop that can be connected to the rest of the geometry. We consider a regularized version of the LQG Hamiltonian and focus on the Euclidean term $EEF[A]$ with arbitrary Immirzi-Barbero parameter, which leads to a generalized ansatz for the Hamiltonian of the two-legged single-loop. This ansatz has the surprising property that the areas carried by the two bulk links can be regrouped in total and relative areas with decoupled dynamics, similar to the decomposition of the 2-body mechanical problem in terms of center-of-mass and relative motion. Both total and relative areas individually satisfy a second order differential equation, which we identified as the cubic non-linear Schr\"odinger equation.

Assuming that the boundary links carry fixed areas, we obtain an oscillatory evolution of the relative area, which we identify as the bounded modes, and a faster-than-exponential evolution of the total area, which can be thought of as scattering modes. The latter case actually displays an explosive growth, which always diverges in finite time. This signals a curvature singularity, i.e.~UV modes that probably ought to be renormalized or tamed down by modifying the Hamiltonian. Nonetheless, these divergences can be pushed to infinite time by natural field-dependent redefinition of the lapse function, as standard in general relativity. This relates to the discussion on the definitions and choices of observers and clocks in (loop) quantum gravity.
We interpret this simple analytically-solvable system of the two-legged single-loop as the equivalent of the harmonic oscillator for LQG, illustrating the basic features of the dynamics of the theory.

Finally, we discussed how to extend the model to time-dependent and dynamical boundary conditions, in order to take into account the impact of the evolution of the exterior geometry on this elementary building block in a more realistic and credible way.
Following this line of thought, we envision few directions of development and refinement for the investigation and technical improvements of the candy graph dynamics in the short term:
\begin{itemize}
\item We could study more thoroughly the dynamics generated by exchange terms between bulk areas/spins and boundary areas/spins, and search for analytical solutions, to allow for dynamical boundary states, bulk-boundary interactions and scalability (i.e. realistic gluing of candy graphs into a spin network).

\item We should study the dynamics generated by taking into account the Lorentzian term in the Hamiltonian, and more generally interaction terms between nodes, such as the  volume-volume couplings (as envisioned e.g. in \cite{Feller:2015yta}). This would allow to analyze the relevance of the 3-curvature potential term in the ADM and LQG Hamiltonians in the microscopic dynamics of the geometry. As we explained earlier in this paper, those terms do not a priori affect the dynamics of the areas, which we derived here, but they will definitely affect the dynamics of the angular direction of the fluxes, as well as generate non-trivial couplings between nodes of the graph.

\item As  discussed in section \ref{sec:architecture}, the purpose of understanding perfectly the dynamics of a single candy graph in LQG is similar to understanding the spectrum of an atom in chemistry: we would like to put them together and study their emergent collective dynamics, for instance the possibility of wave propagation and dissipation in chains or lattices of candy graphs (as drawn on fig.\ref{fig:coilchain}), or the renormalization flow under coarse-graining of melonic imbrication of candy graphs (as drawn on fig.\ref{fig:melonic}).

\item  Looking at the candy graph as a 2-vertex building block of spin networks, we proposed to consider it as a basic template of the LQG dynamics, similarly to the 2-body problem in (Newtonian) gravitational physics. It is then natural to look beyond the 2-vertex case and consider more elaborated building blocks of the quantum geometry in the search for new phenomenology for LQG. Would increasing the number of vertices lead to a richer set of modes and excitations than the present study? For instance, would 3-vertex configurations reveal a richer physics like the gravitational 3-body problem with chaotic behavior, subtler stability and Lagrange points?
\end{itemize}

Then in the long run, we would like to go beyond the technical investigation, along three axes.
First, at the classical level, we wish to compare the LQG Hamiltonian ansatz with the renewed insights on conserved charges and symplectic flux balance equations for general relativity, see e.g. \cite{Donnelly:2020xgu,Fiorucci:2021pha}. For instance, an advanced consistency check would be to understand if the candy graph kinematics and Hamiltonian carry a representation of the algebra of boundary charges, either for finite-distance boundaries or asymptotically.
Then, we want to move on to the quantum theory and investigate purely quantum features of the quantum candy graphs, such as the entanglement build-up between the two spin network nodes. We have a dual strategy in mind: on the one hand, study the spectrum of the Hamiltonian, and on the other hand, study the effective evolution of volume, area and curvature expectation values and their higher moments and the corresponding quantum uncertainty.
Third, the necessity of considering graph-changing dynamics in loop quantum gravity has often been emphasized (see for instance \cite{Assanioussi:2018zit, Assanioussi:2020fsz, Guedes:2024zbu,Guedes:2024duc}). This is what clearly distinguishes LQG from lattice approaches to general relativity. Nevertheless, a possibility of studying varying graph dynamics while staying on a given background has already been proposed in \cite{Lewandowski:2014hza, Assanioussi:2015gka, Charles:2016xwc} and consist in allowing for ``little loops'' to be created and annihilated at every node, representing local curvature excitations developing on top of quanta of volumes at each node and not only between nodes, see for instance \cite{Pranzetti:2012dd,Anza:2017dkd}. The general idea is to consider little loops at both nodes of the candy graphs, allow for spin exchange between them, the bulk links and the boundary links, study the resulting dynamics, the growth of the number of little loops versus their spins and the resulting evolution of the entanglement and correlations between the two nodes.

From a broader perspective, these questions lead to deeper considerations about the renormalization flow of loop quantum gravity. Indeed, although spin networks are the eigenstates of geometry at the Planck scale, they might not be the most relevant structure at a more coarse-grained level at larger scales. Diving in the yet unknown phase diagram of the theory, one can imagine glassy phases or liquid phases, following the intuition of the hydrodynamical-like reformulations of general relativity, with flexible connectivity between spin network nodes, similarly to the quantum condensate proposal of group field theory (see e.g. \cite{Carrozza:2013oiy,Gielen:2016dss,Oriti:2021oux}). Choosing which phase to study should determine the relevant parameters and perturbative expansions for the LQG dynamics. For instance, should the relevant parameter to organize spin network states be the number of bulk nodes (similarly to atomic physics) or bulk loops (like Feynman diagrams in quantum field theory) or boundary edges (in a holographic perspective) or something more subtle like turbulent cascades with energy flowing from IR modes to UV modes? Better understanding these questions will shed light on the relevant spin network architecture depending on the considered scales of energy and length.

Finally, since our evolution equations are the same non-linear equations that drive Josephson junctions, let us conclude with the enticing perspective of simulating glued-candy-graphs architecture using Josephson junction array and thereby implementing analogue loop quantum gravity systems in a direct way on quantum circuits.

%%%%%%%%
\section*{Acknowledgement}
%%%%%%%%
This research was funded in part by the National Science Centre, Poland, through SONATA 19 grant no.~2023/51/D/ST2/00296. For the purpose of Open Access, the authors have applied a CC-BY public copyright licence to any Author Accepted Manuscript (AAM) version arising from this submission.

%%%%%%%%
%\appendix
%%%%%%%%

%%%%%%%%%%%%%%%%%%
%%%%%%%%%%%%%%%%%%

%%%%%%%%%%
% BIBLIOGRAPHY
%%%%%%%%%%
\newpage
\bibliographystyle{bib-style}
\bibliography{2VertexDyn.bib}

\end{document}